\documentclass[letterpaper,11pt]{article}
\usepackage[x11names]{xcolor}
\usepackage{colortbl} 
\usepackage{amsmath}
\usepackage{amssymb}
\usepackage{amsfonts}
\usepackage{graphicx} 
\usepackage{ textcomp }
\usepackage{ dsfont }
\usepackage{subcaption}
\usepackage{mathbbol}
\usepackage{mathrsfs}
\usepackage[margin=1in,letterpaper]{geometry} 
\usepackage[final]{hyperref} 
\hypersetup{
	colorlinks=true,       
	linkcolor=blue,        
	citecolor=blue,        
	filecolor=magenta,     
	urlcolor=blue         
}
\newcommand\blfootnote[1]{%
  \begingroup
  \renewcommand\thefootnote{}\footnote{#1}%
  \addtocounter{footnote}{-1}%
  \endgroup
}

\begin{document}

\title{Gauge boson masses from the scale-dependent interplay between the gravitational and the electroweak sector}
\author{
Cristobal Laporte $^{\spadesuit}$
\blfootnote{$^{\spadesuit}$ email:
\href{mailto:cristobal.laportemunoz@ru.nl}
{\nolinkurl{cristobal.laportemunoz@ru.nl}}
},
Benjamin Koch $^{\diamondsuit, \triangle}$
\blfootnote{$^{\diamondsuit}$ email:
\href{mailto:benjamin.koch@tuwien.ac.at}
{\nolinkurl{benjamin.koch@tuwien.ac.at}}
},
Felipe Canales $^{\triangle}$
\blfootnote{$^{\triangle}$ email:
\href{mailto:facanales@uc.cl}
{\nolinkurl{facanales@uc.cl}}
},
\'Angel Rinc\'on $^{\clubsuit}$ 
\blfootnote{$^{\clubsuit}$email:
\href{mailto:aerinconr@academicos.uta.cl}
{\nolinkurl{aerinconr@academicos.uta.cl}}
}
\\
\small{$^{\triangle}$ Instituto de Física, Pontificia Universidad Católica de Chile, Casilla 306, Santiago, Chile}
\\
\small{
$^{\diamondsuit}$ Institut fur Theoretische Physik, Technische Universitat Wien, }\\
\small{Wiedner Hauptstrasse 8-10, A-1040 Vienna, Austria}
\\
\small{
$^{\spadesuit}$ Institute for Mathematics, Astrophysics and Particle Physics (IMAPP), Radboud University,}\\
\small{Heyendaalseweg 135, 6525 AJ Nijmegen, The Netherlands
}
\\
\small{
$^{\clubsuit}$ Sede Esmeralda, Universidad de Tarapac\'a,
Avda. Luis Emilio Recabarren 2477, Iquique, Chile.
}
}
\maketitle

\begin{abstract}
In this paper we explore the interplay of scale dependence arising
in the electroweak and the gravitational sector.
Using the method of variational parameter setting leads
naturally to spontaneous symmetry breaking and thus to electroweak gauge boson masses.
We explore the phenomenological prospect of this model and the mapping
to the program asymptotic safety.
\end{abstract}

\tableofcontents

\section{Introduction}

Spontaneous Symmetry Breaking (SSB) is an essential concept of theoretical physics.
Numerous phenomena, such as the spatial order of ferromagnets~\cite{Landau:1961ha,Sadler:2006to}, convecting fluids~\cite{Velarde:1980go,Sneddon:1981lu}, and superconductivity~\cite{Ginzburg:1950sr,Weinberg:1986cq} are described by SSB. In particle physics, this mechanism allowed for a successful description
of gauge boson masses.
This lead to what is now known as the Electroweak Standard Model (EW-SM) of particle physics~\cite{Englert:1964et,Higgs:1964pj,Glashow:1967rx,Weinberg:1967tq}.
However, the interactions, like the ones arising from the quartic term in the Higgs potential,
seem to come at the price of an unnatural tuning of the parameters 
of the model~\cite{tHooft:1979rat,deGouvea:2014xba,Georgi:1974yf,Susskind:1978ms,Dine:2015xga,Giudice:2008bi}.

Even though SSB is the most accepted mechanism to generate gauge boson masses, there are
alternative descriptions.
Some of these alternatives are the  
St\"uckelberg mechanism~\cite{Stueckelberg:1938hvi} (only works for Abelian gauge symmetry), 
an asymptotically safe model for gauge boson masses (does not have dynamical Higgs~ \cite{Calmet:2010ze}), 
composite Higgs sectors \cite{Kaplan:1983sm,Dugan:1984hq},  
composite gauge boson sectors~\cite{Abbott:1981re}, 
or even extra dimensions~\cite{Csaki:2003zu}.
While the first two can not serve as a description of the observed 
massive $\text{SU(2)}$ gauge bosons with a dynamical Higgs boson, 
the latter can provide these phenomenological features at the price of adding
new, yet undiscovered, degrees of freedom.

The idea of this paper is to implement Higgs-like SSB without a quartic coupling and the need for new particle sectors. 
This is achieved by considering the gravitational sector and
quantum-induced scale-dependence (SD). Quantum scale-dependence is known and familiar for all effective quantum field theories, but it is also a likely feature the gravitational sector~\cite{Rosten:2010vm,Reuter:2007rv,Pawlowski:2018ixd,Shaposhnikov:2009pv,Eichhorn:2017als,Kwapisz:2019wrl,Wetterich:2019rsn}.
Scale-dependence is well known and understood in the particle sector 
\cite{Altaisky:2020dft,Wu:2013ei,ALICE:2021nir,Domdey:2009bg,Deur:2016tte,Wu:2014iba}
and also well known
in the gravitational sector \cite{Contreras:2013hua,Koch:2015nva,Koch:2016uso,Rincon:2018lyd,Canales:2018tbn,Alvarez:2020xmk}, but only recently considered simultaneously in both sectors \cite{Koch:2020baj}. 
Here, we consider this simultaneous SD in the context of variational parameter setting~\cite{Koch:2020baj}.
It is found that SSB occurs only after considering all these elements and after setting the arbitrary renormalization scale $k$ to the optimal scale $k_{opt}$.
It has thus certain similarity to  the Coleman-Weinberg mechanism, where SSB also occurs after
choosing the renormalization scale for radiative corrections~\cite{Coleman:1973jx}.

This paper generalizes a preceding work on the Abelian Higgs mechanism~\cite{Koch:2020baj}. Thus, it is
organized as follows: after this brief introduction, we discuss, in Sec. II, the method used along with the manuscript, reviewing the SSB in the electroweak sector, the scale setting, and the optimal effective action. Then, in section III, we show the main results: i) the gauge-invariant in the optimal action, 
ii) the expression for such action (detailing the corresponding expansion of the parameters involved), 
iii) the solution of the $\beta$ functions, 
and 
iv) the mass and vacuum expectations values of scalar and gauge fields. 
A comparison with the functional renormalization group is performed in section IV. Finally, we conclude in the last section and elaborate on future work ideas.

\section{Method}\label{Method}

\subsection{SSB in electroweak sector}

In particle physics, the Lagrangian that describes the electroweak interactions is invariant under $\text{SU(2)}\otimes\text{U(1)}$ gauge transformations. 
Furthermore, gauge invariance is intimately linked to spontaneous electroweak symmetry breaking and, thus, generating massive gauge bosons from massless gauge fields. 
Let us consider the electroweak sector of the standard model represented by the following Lagrangian density,
\begin{equation}
    \mathcal{L_{\text{EW}}}=-\frac{1}{4}B_{\mu\nu}B^{\mu\nu}-\frac{1}{4}W^a_{\mu\nu}W^{a,\mu\nu}+(D_\mu H)^\dagger(D^\mu H)+m^2H^\dagger H-\lambda(H^\dagger H)^2,
\end{equation}
where the covariant derivative is given by,
\begin{equation}
    D_\mu H=\partial_\mu H-\frac{i}{2}gW_\mu H+\frac{i}{2}g'B_\mu H.
\end{equation}
Expanding the
Higgs field around its ground state, with a vacuum expectation value $v=\frac{m}{\sqrt{\lambda}}$, and the gauge is set to configure the field,
one can write
\begin{equation}
    H=\frac{1}{\sqrt{2}}\begin{pmatrix}0\\v+h\end{pmatrix}.
\end{equation}
This choice breaks the $\text{SU(2)}$ symmetry in the Lagrangian, while the invariance under the local transformations U(1) remains intact.
The kinetic term of the Higgs field reads
\begin{align}\label{eq_KinH0}
    (D_\mu H)^\dagger(D^\mu H)=&\frac{g^2}{4}(W^1_\mu+iW^2_\mu)(W^{1,\mu}-iW^{2,\mu})\left(\frac{v}{\sqrt{2}}+\frac{h}{\sqrt{2}}\right)^2+\frac{1}{2}\partial_\mu h\partial^\mu h\nonumber\\
    &+\frac{1}{4}(gW^3_\mu+g'B_\mu)(gW^{3,\mu}+g'B^\mu)\left(\frac{v}{\sqrt{2}}
+\frac{h}{\sqrt{2}}
\right)^2
\end{align}
%
Performing the following field redefinitions, 
\begin{align}
    W_\mu=&\frac{1}{\sqrt{2}}(W^1_\mu+iW^2_\mu)\\
    Z_\mu=&\frac{gW^3_\mu+g'B_\mu}{\sqrt{g^2+g'^2}}\\
    A_\mu=&\frac{-g'W^3_\mu+gB_\mu}{\sqrt{g^2+g'^2}},
\end{align}
%
one can express (\ref{eq_KinH0})  as, 
\begin{equation}\label{StandardCovDer}
    (D_\mu H)^\dagger(D^\mu H)=\frac{1}{2}\partial_\mu h\partial^\mu h+\frac{g^2v^2}{4}W^\dagger_\mu W^{\mu}+\frac{v^2(g^2+g'^2)}{8}Z_\mu Z^\mu+\text{higher order terms}.
\end{equation}
%
%
Thus, the symmetry breaking induced the masses $m_W=\displaystyle\frac{gv}{2}$ and $m_Z=\displaystyle\frac{v}{2}\sqrt{\frac{g^2+g'^2}{2}}$ for the gauge bosons W and Z, while the $\text{U(1)}_\text{em}$ gauge boson remained massles $m_\gamma=0$.

\subsection{Scale setting and optimal effective action}
\label{subsec_scaleset}

This subsection briefly introduces the essential ingredients and main ideas behind the scale-setting procedure and the resulting optimal (effective) action.
Let us start by considering generic effective action labelled as $\Gamma_k[\phi, A_\mu, \psi, g_{\mu\nu}, \cdots]$. 
This action is characterized by the field content $(\phi, A_\mu, \psi, g_{\mu \nu})$, 
the scale-dependent (SD) couplings $\alpha_{i}(k)$ between the fields, and the renormalization scale $k$.
This scale plays a crucial role in this approach.
On the one hand, the renormalization group (RG) equations determine the SD of the couplings $\alpha_{i}(k)$
and while, on the other hand, these couplings appear in the observables of the theory.
To obtain a prediction for these observables, one has to set the RG scale $k$ in terms of the physical parameters $P_i$ of the system under consideration. This procedure is called scale setting. 
An ad-hoc choice $k=k(P_i)$ induces a subtle degree of arbitrariness to the value of the observable prediction of the theory.
Due to the typical logarithmic dependence on $k$, this arbitrariness is part of the very small systematic uncertainty. Thus, one can typically choose the scale in a bold way, for example, of the order
of the invariant momentum square or invariant momentum square of a collision, without affecting the numerical value of the observable. Such a bold choice is, however, not always satisfactory. There are various examples where the scale setting significantly impacts the physical meaning of a given model.
\begin{itemize}
\item[a)] High precision observables,\\
can be sensitive to different (IR) cut-off scales and thus
to the choices of the energy scale $k$. This means that physical parameters, such as the quark masses, must be treated with care since they depend on this choice~\cite{Hoang:2018zrp}.
\item[b)] Poor convergence of perturbation theory,\\
as is the case for QCD approaching the $\Lambda_{\text{QCD}}$ scale,
it comes with increased sensitivity to the scale setting $k$.
To mitigate this limitation by formulating scale-setting prescriptions
which makes observables more insensitive to this choice.
This is the philosophy of the ``Callan-Symanzik'' equations~\cite{Callan:1970yg,Symanzik:1970rt},
the ``principal of minimal sensitivity~\cite{Stevenson:1981vj}, and of
 the ``principle of maximal conformality''~\cite{Brodsky:1982gc,Brodsky:2011ig}.
\item[c)]  The dimensionful couplings,\\
as they appear in an effective description of quantum gravity, imply a strong 
power law dependence on $k^2$. To arrive at reliable predictions of
such a theory, one should seek a scale-setting prescription that mitigates this strong dependence.
\end{itemize}
In this paper, we are primarily concerned with scenario c), where strong SD arises from the gravitation sector of the theory, but it is coupled to the particle sector of the theory (scenario a)), which can rely on precise experimental data.
We will thus apply a variational scale setting method, which has to be particularly successful in taming for strong SD effects of scenario c). This method, the variational parameter setting~\cite{Koch:2010nn,Domazet:2012tw,Koch:2014joa},
applies a variational principle the effective action $\Gamma_k$,
\begin{equation}\label{variational_principle}
    \frac{\delta\Gamma_k}{\delta k}\Bigl|_{k=k_{\text{opt}}}=0.
\end{equation}
This condition is imposed to minimize the otherwise enormous arbitrariness of the scale setting while preserving 
the local and global symmetries of the original Lagrangian.
A scale ,which is optimal in this sense, is obtained by solving \eqref{variational_principle} for $k_{\text{opt}}$.
This optimal scale $k_\text{opt}$ can be replaced in the effective action to obtain an optimal action that does not depend on the scale, only depend on the physical variable (fields and couplings).
Other scale setting approaches such as the improving solution approach, the use
of proper dimensionful invariants, or the use of approximated conservation laws
\cite{Uehling:1935uj,Dittrich:1985yb,Bonanno:1998ye,Wetterich:1992yh,Reuter:1993kw,Reuter:1993kw,Reuter:1992uk,Bonanno:2000ep,Litim:2007iu,Bertini:2019xws,Platania:2020lqb} are typically only applicable when 
a specific classical solution is known to describe the physical system under consideration reasonably well.

\section{Results and Analysis}


\subsection{Gauge Invariance in the Optimal Effective Action: the U(1) case}\label{GaugeInv}

To the mechanism recently introduced in \cite{Koch:2020baj} and to this work, the gauge invariance is important not only for the initial Lagrangian describes interaction but also for the effective action with running couplings and the optimized action. The U(1) Lagrangian without quartic interaction discussed in \cite{Koch:2020baj} is
\begin{equation}
\Gamma_{k}=\int d^4x \sqrt{-g}\left\{\kappa_k(2\Lambda_{k}-R)+\frac{m^2_{k}}{2}\phi^{*}\phi \right.
        \left. +\frac{1}{2}(D_{\mu}\phi)^{*}(D^{\mu}\phi)-\frac{1}{4e^2_{k}}F^{\mu\nu}F_{\mu\nu}\right\}.
\end{equation}
This action presents a gauge symmetry. The choice of the optimal scale
\begin{equation}\label{k_opt_U(1)}
    k_{\text{opt}}=-\frac{8\xi_{\tilde{\Lambda},1}+4\xi_{m,1}|\phi|^2 -\xi_{e,1}F^{\mu\nu}F_{\mu\nu}}{2\left(8\xi_{\tilde{\Lambda},2}+4\xi_{m,2}|\phi|^2-\xi_{e,2}F^{\mu\nu}F_{\mu\nu}\right)},
\end{equation}
only contains gauge invariant quantities
and the parameter expansions $\xi_{\tilde{\Lambda},2}$, $\xi_{\tilde{\Lambda},2}$, $\xi_{m,1}$, $\xi_{m,2}$, $\xi_{e,1}$ and $\xi_{e,2}$ are defined in \cite{Koch:2020baj}. Therefore, when replacing $k_\text{opt}$ in the effective action $\Gamma_k$, we proved in the appended section~\ref{AppendixB} that the optimized action also presents invariance under the U(1) transformations: $\phi\rightarrow \phi+i\alpha(x)\phi$ and $A_\mu\rightarrow A_\mu-\partial_\mu\alpha(x)$,
\begin{align}
    \Gamma_{\text{opt}}=&\int d^4x\sqrt{-g}\Big[2\Lambda_0+\left(m^2_0+A_\alpha A^\alpha\right)|\phi|^2+iA^\alpha\left(\phi\partial_\alpha\phi^*-\phi^*\partial_\alpha\phi\right)+\partial_\alpha\phi^*\partial^\alpha\phi-\frac{1}{4e^2_0}F^{\mu\nu}F_{\mu\nu}\nonumber\\
    &+\frac{k^2_{\text{opt}}}{2}\Big\{4\xi_{\tilde{\Lambda},2}+2\xi_{m,2}|\phi|^2+4\xi_{\tilde{\Lambda},1}\left(8\xi_{\tilde{\Lambda},2}+4\xi_{m,2}|\phi|^2-\xi_{e,2}F^{\mu\nu}F_{\mu\nu}\right)-\frac{\xi_{e,2}}{2}F^{\mu\nu}F_{\mu\nu}\nonumber\\
    &+\xi_{e,1}\left(4\xi_{\tilde{\Lambda},2}+2\xi_{m,2}|\phi|^2-\xi_{e,2}F^{\mu\nu}F_{\mu\nu}\right)+4\xi_{m,1}|\phi|^2\left(4\xi_{\tilde{\Lambda},2}+2\xi_{m,2}|\phi|^2-\xi_{e,2}F^{\mu\nu}F_{\mu\nu}\right)\Big\}\Big],
\end{align}
Introducing the optimal scale into the action does not alter its gauge invariance.
The same is true for $\text{SU(2)}$.

%

\subsection{Optimal Effective Action: the $\text{SU(2)}\times\text{U(1)}$ case}

In this section, the methodology discussed in section~\eqref{Method} will be applied to a Lagrangian
with the gauge symmetry the symmetry  $\text{SU(2)}\times\text{U(1)}$. The breaking of this symmetry 
$\text{SU(2)}\times\text{U(1)} \rightarrow \text{U(1)}_\text{em}$ will only occur
at the final step, after all quantum effects and scale setting procedures have been implemented. 

The model used in this work considers a minimal coupling between the gravitational sector, an Einstein-Hilbert action, and a matter sector.
The latter contains a 
complex doublet Higgs multiplet $H$ with hypercharge $\frac{1}{2}$,
a  hypercharge gauge boson field $B_{\mu}$, with $B_{\mu\nu}=\partial_\mu B_\nu - \partial_\nu B_\mu$,
and a $\text{SU(2)}$ gauge boson $W_\mu^a$ with the corresponding field strength $W^a_{\mu\nu}=\partial_{\mu}W^a_{\nu}-\partial_{\nu}W^a_{\mu}-g\epsilon^{abc}W^b_{\mu}W^c_{\nu}$,
\begin{equation}\label{Weakaction}
    \Gamma_k = \int d^4x \sqrt{-g} \left\{\frac{-R+2\Lambda_k}{16\pi G_k} + m^2_k H^\dagger H - \frac{1}{4}W^a_{\mu\nu}W^{a,\mu\nu} -\frac{1}{4}B_{\mu\nu}B^{\mu\nu} + (D_{k,\mu} H)^\dagger(D^\mu_{k} H)\right\}.
\end{equation}
Here, the scale-dependent covariant derivative is defined as, 
\begin{equation}\label{CovDer}
    D_{k,\mu}H \equiv \partial_\mu H+\frac{1}{2}ig_kW^a_\mu \sigma^a H - \frac{1}{2}ig^{'}_kB_\mu H,
\end{equation}
where $g_k$ and $g^{'}_k$ are the $\text{SU(2)}$ and $\text{U(1)}$ scale-dependent couplings, respectively. 
The matrices $\sigma^a$ are the canonically normalized $\text{SU(2)}$ generators. The Higgs doublet $H$ and the SU(2) generator, written in matrix form, are defined by 
\begin{align}\label{FieldStrenght}
    H &= \begin{pmatrix}\phi_1\\\phi_2\end{pmatrix} \\
    W^a_\mu \, \sigma^a
    &= \begin{pmatrix}\frac{W_{3,\mu}}{2} && \frac{W_{1,\mu}-iW_{2,\mu}}{2}\\\frac{W_{1,\mu}+iW_{2,\mu}}{2} && -\frac{W_{3,\mu}}{2}\end{pmatrix},
\end{align}
respectively. Plugging the field definitions~(\ref{FieldStrenght}) in the kinetic term for the Higgs field, one finds explicitly
\begin{align}
    (D_{k,\mu} H)^\dagger(D^\mu_{k} H) =& \,\frac{g_kg'_k}{2}B^\alpha\left(-W^1_\alpha(\phi^*_1\phi_2+\phi_1\phi^*_2) + \mathrm{i}W^2_\alpha(\phi^*_1\phi_2-\phi_1\phi^*_2)+W^3_\alpha(|\phi_2|^2-|\phi_1|^2)\right) \nonumber\\
    &+\frac{ig_kW^{3,\alpha}}{2}\left(\phi_1\partial_\alpha\phi^*_1-\phi^*_1\partial_\alpha\phi_1-\phi_2\partial_\alpha\phi^*_2+\phi^*_2\partial_\alpha\phi_2\right)+\frac{1}{4} (g'^2_{k}\tilde{B}^2+g^2_k\tilde{W}^2)\tilde{\phi}^2\nonumber\\
    &+ \partial_\mu H^\dagger \partial^\mu H +\frac{\mathrm{i} g'_k}{2}B^\alpha\left(\phi^*_1(\partial_\mu\phi_1)-\phi_1(\partial_\mu\phi^*_1)+\phi^*_2(\partial_\mu\phi_2)-\phi_2(\partial_\mu\phi^*_2)\right) \nonumber\\
    &+\frac{\mathrm{i}g_k}{2}W^{1,\alpha}\left(\phi_2\partial_\alpha\phi^*_1-\phi^*_2\partial_\alpha\phi_1+\phi_1\partial_\alpha\phi^*_2-\phi^*_1\partial_\alpha\phi_2\right)\nonumber\\
    &+\frac{\mathrm{i}g_k}{2}W^{2,\alpha}\left(\phi^*_2\partial_\alpha\phi_1-\phi_2\partial_\alpha\phi^*_1+\phi^*_1\partial_\alpha\phi_2-\phi_1\partial_\alpha\phi^*_2\right),
\end{align}
where $\tilde{W}^2\equiv W^2_1+W^2_2+W^2_3$, $\tilde{B}^2\equiv B^\alpha B_\alpha$ and $\tilde{\phi}^2\equiv \phi^2_1+\phi^2_2$. Since we are interested in exploring whether this model can generate the observed masses for the gauge bosons in the electro-weak sector, the expansion of the gauge couplings will be done in the infrared (IR). To deal with the logarithmic divergences appearing in the electro-weak couplings in the deep infrared scale $k\rightarrow 0$, the RG scale is split into its fixed IR reference part $k_0=m_0$ and a variable part $k'$ as $k=m_0+k'$. The couplings can now be expanded in the vicinity of the reference scale $m_0$, using the dimensionless quantity $\frac{k'}{m_0}$ as the expansion parameter,
\begin{subequations}\label{CouplingExpansion}
\begin{align}
    \frac{1}{G_k} &= \frac{1}{G_0} + \xi_{G,1}\left(\frac{k'}{m_0}\right) + \xi_{G,2}\left(\frac{k'}{m_0}\right)^2 +     \mathcal{O}\left(\frac{k'}{m_0}\right)^3,\\
    \label{GammaTildeExpanssion}\tilde{\Lambda}_k &= \tilde{\Lambda}_0 + \xi_{\tilde{\Lambda},1}\left(\frac{k'}{m_0}\right) + \xi_{\tilde{\Lambda},2}\left(\frac{k'}{m_0}\right)^2 + \mathcal{O}\left(\frac{k'}{m_0}\right)^3,\\
    m^2_k &= m^2_0 + \xi_{m,1}\left(\frac{k'}{m_0}\right) + \xi_{m,2}\left(\frac{k'}{m_0}\right)^2 + \mathcal{O}\left(\frac{k'}{m_0}\right)^3,\\
    g'_k &= g'_0 + \xi_{g',1}\left(\frac{k'}{m_0}\right) + \xi_{g',2}\left(\frac{k'}{m_0}\right)^2 + \mathcal{O}\left(\frac{k'}{m_0}\right)^3,\\
    g_k &= g_0 + \xi_{g,1}\left(\frac{k'}{m_0}\right) + \xi_{g,2}\left(\frac{k'}{m_0}\right)^2 + \mathcal{O}\left(\frac{k'}{m_0}\right)^3.
\end{align}
\end{subequations}
The set of coefficients $\xi_{i,j}$ (with $i=m,g',g$ and $j=1,2$) can be obtained by integrating the corresponding beta functions of weak interactions, while the gravitational coefficient will remain arbitrary. In the set~(\ref{CouplingExpansion}) we have introduced the variable $\tilde{\Lambda}$ defined by $\tilde{\Lambda}_k\equiv \frac{\Lambda_k}{G_k}$. 
Now, the scale-setting procedure described in subsection \ref{subsec_scaleset}
can be applied to the effective action (\ref{Weakaction}) with the expanded couplings (\ref{CouplingExpansion})
by imposing the variational condition (\ref{variational_principle}).
The solution of this condition defines the optimal scale $k_{\text{opt}}$. This prescription allows resolving the renormalization-point ambiguity by selecting a single scale as a function of the dynamical variables of the problem, 
\begin{equation}\label{OptimalScaleComplete}
    k_{\text{opt}}=\frac{\mathcal{N}}{\mathcal{D}}.
\end{equation}
Here,
\begin{eqnarray}
    \mathcal{N}=&4\xi_{\tilde{\Lambda},1}+(g_0\xi_{g',1}+g'_0\xi_{g,1})\left[-W^1_\alpha B^\alpha (\phi^*_1\phi_2+\phi_1\phi^*_2)+\mathrm{i}W^2_\alpha B^\alpha (\phi^*_1\phi_2-\phi_1\phi^*_2)-W^3_\alpha B^\alpha (|\phi_1|^2-|\phi_2|^2)\right]\nonumber\\
    &+g'_0\xi_{g',1}\tilde{B}^2\tilde{\phi}^2+g_0\xi_{g,1}\tilde{W}^2\tilde{\phi}^2+2 \xi_{m,1}\tilde{\phi}^2+\mathrm{i}\xi_{g',1} B^\alpha (\phi^*_1 \partial_\alpha \phi_1-\phi_1\partial_\alpha\phi^*_1+\phi^*_2\partial_\alpha\phi_2-\phi_2\partial_\alpha\phi^*_2)\nonumber\\
    &-\mathrm{i}\xi_{g,1}W^{1,\alpha}(\phi^*_1\partial_{\alpha}\phi_2-\phi_1\partial_\alpha\phi^*_2+\phi^*_2\partial_\alpha\phi_1-\phi_2\partial_\alpha\phi^*_1)\nonumber\\
    &+\xi_{g,1}W^{2,\alpha}(\phi^*_2\partial_\alpha\phi_1-\phi^*_1\partial_\alpha\phi_2+\phi_2\partial_\alpha\phi^*_1-\phi_1\partial_\alpha\phi^*_2)\nonumber\\
    &-\mathrm{i}\xi_{g,1}W^{3,\alpha}(\phi^*_1\partial_\alpha\phi_1-\phi_1\partial_\alpha\phi^*_1+\phi_2\partial_\alpha\phi^*_2-\phi^*_2\partial_\alpha\phi_2)
\end{eqnarray}
and the denominator of (\ref{OptimalScaleComplete}) can be written as,
\begin{eqnarray}
    \mathcal{D} = \mathbb{D}_1 + \mathbb{D}_2 \, (g_0\,\xi_{g',2} + \xi_{g,1}\,\xi_{g',1} + g'_0\,\xi_{g,2}) + \mathbb{D}_3 \, \xi_{g',2} + \mathbb{D}_4 \, \xi_{g,2},
\end{eqnarray}
where we have defined the quantities,
\begin{subequations}
\begin{align}
    \mathbb{D}_1 &= 8 \, \xi_{\tilde{\Lambda},2} + 4 \, \xi_{m,2} \tilde{\phi}^2 + \xi^2_{g,1}\tilde{W}^2\tilde{\phi}^2 + \xi^2_{g',1}\tilde{B}^2\tilde{\phi}^2 + 2\,(\xi_{g,2}\, g_0\tilde{W}^2+\xi_{g',2}\, g'_0\tilde{B}^2)\,\tilde{\phi}^2,\\
    \mathbb{D}_2 &= -2 \, B^\alpha \, \left[W^1_\alpha(\phi_1\phi^*_2 + \phi^*_1\phi_2) - \mathrm{i}W^2_\alpha (\phi^*_1\phi_2-\phi_1\phi^*_2) + W^3_\alpha (|\phi_1|^2-|\phi_2|^2)\right],\\
    \mathbb{D}_3 &= -2 \, \mathrm{i} \, B^\alpha \left[(\partial_\alpha\phi^*_1)\phi_1 - (\partial_\alpha\phi_1)\phi^*_1 + (\partial_\alpha\phi^*_2)\phi_2 - (\partial_\alpha\phi_2)\phi^*_2\right]\\
    \mathbb{D}_4 &= 2\,\mathrm{i}\left[(W^{1,\alpha}\partial_\alpha\phi^*_2-\mathrm{i}W^{2,\alpha}\partial_\alpha\phi^*_2+W^{3,\alpha}\partial_\alpha\phi^*_1)\phi_1 - (W^{1,\alpha}\partial_\alpha\phi_2+\mathrm{i}W^{2,\alpha}\partial_\alpha\phi_2+W^{3,\alpha}\partial_\alpha\phi_1)\phi^*_1\right]\nonumber\\
    &+2\,\mathrm{i}\left[(W^{1,\alpha}\partial_\alpha\phi^*_1+\mathrm{i}W^{2,\alpha}\partial_\alpha\phi^*_1-W^{3,\alpha}\partial_\alpha\phi^*_2)\phi_2 - (W^{1,\alpha}\partial_\alpha\phi_1-\mathrm{i}W^{2,\alpha}\partial_\alpha\phi_1-W^{3,\alpha}\partial_\alpha\phi_2)\phi^*_2\right].
\end{align}
\end{subequations}
When the optimal scale~(\ref{OptimalScaleComplete}) is inserted back into the effective action~(\ref{Weakaction}), one gets the optimal $k$-independent effective action which posses the following structure,
\begin{align}\label{OptimalEA}
    \Gamma_{\text{opt}} = \int d^4x\sqrt{-g} \left\{+\frac{\mu^2}{2}|H|^2-\frac{\lambda}{4}|H|^4+\mathcal{L}_{\text{kin}}+\mathcal{L}_{\text{const}}+\mathcal{L}_{\text{coup}}+\mathcal{L}\left(W,B\right)+\mathcal{O}(H^\dagger H)^3\right\},
\end{align}
where $\mathcal{L}_{\text{kin}}$ contains kinetic terms for the real components of the Higgs and gauge fields. Couplings with higher-order factors and Ricci scalar quantities are collected into $\mathcal{L}_{\text{coup}}$, and the Lagrangian part, which is independent of the Ricci scalar, electroweak strength, and scalar field are named $\mathcal{L}_{\text{const}}$. In~(\ref{OptimalEA}), the effective potential associated with the scalar sector has been expanded to order $(H^\dagger H)^2$ in a weak field approximation. At this point, several comments are in order:
\begin{enumerate}
    \item The optimal action~(\ref{Weakaction}) does not have a quartic self-interaction $(H^\dagger H)^2$, which means 
    that the Higgs mechanism cannot take place to generate masses for the Higgs, W, and Z bosons. 
    However, introducing the gravity sector, particularly a vacuum energy-density term, induces a Higgs-like SSB scalar potential once the VPS is applied to replace the energy scale as a function of dynamical variables. Additionally, the $\mathcal{L}\left(W,B\right)$ part includes the different interactions between the gauge fields $\text{SU(2)}$ and $\text{U(1)}_\text{Y}$, thus containing the necessary ingredients for their mass terms. 
    \item Physical observables derived from the optimal action~(\ref{OptimalEA}) retain diffeomorphism invariance. 
    As shown in~\cite{Koch:2010nn,Domazet:2012tw,Koch:2014joa} and section~\ref{GaugeInv}, the VPS maintains all these symmetries present in the Lagrangian~(\ref{Weakaction}). Furthermore, the renormalization scale $k$, as well as other unphysical parameters of the theory, will not be part of the physical quantities derived from the optimal effective action. Another way of circumventing unphysical $k$-dependence is by writing down a RG-improvement solution based on curvature invariants \cite{Held:2021vwd} or similar conditions aiming at background independence~\cite{Pagani:2019vfm}
    \item There are numerous models of gravity-assisted  SSB~\cite{Bekenstein:1986hi,Moniz:1990kt,Krasnov:2011hi,Alexander:2012ge,Rinaldi:2015uvu,Alexander:2016xuy}. These models do, however, require a large value for the cosmological constant, or equivalently, the Ricci curvature. Since the observed value of the cosmological constant is rather low, a straightforward implementation is not realistic, and more sophisticated constructions are required~\cite{George:2015nza}. 
In contrast to this, in the above procedure, the results~(\ref{OptimalScaleComplete}) and~(\ref{OptimalEA}) have been obtained for small Ricci curvature. They are perfectly valid, even in the flat space limit.
    \item If one turns off the SU(2) field and replaces the $U(1)_{Y}$ field and Higgs doublet $H$ by their counterparts $U(1)_{em}$ and Higgs singlet, respectively, we recover the case presented in~\cite{Koch:2020baj}.
    \item The optimal scale~(\ref{OptimalScaleComplete}) and the corresponding optimal action~(\ref{OptimalEA}) 
    are invariant under the  local symmetry SU(2). 
    The Higgs parameters $\mu,\lambda$ in~(\ref{OptimalEA}) are lengthy functions of the  expansion parameters $\xi$'s. To simplify these expressions, one can exploit the invariance under gauge transformations to parametrize the Higgs field as,
    \footnote{Expressing the Higgs field as~(\ref{HiggsParametrization}) means choosing a vacuum that breaks the original symmetry and generate a mass for the corresponding gauge bosons. To do this, the Lagrangian must show SSB, which is why~(\ref{OptimalEA}) is our starting point.}
    \begin{equation}\label{HiggsParametrization}
    H=\text{exp}\left(i\frac{\pi^a\sigma^a}{v}\right)\begin{pmatrix}0\\\phi\end{pmatrix}.
\end{equation}
\end{enumerate}
As in the standard case, it is simpler to study this theory in the unitary gauge, so we set $\pi^a=0$. 
Plugging this gauge choice into the optimal scale~(\ref{OptimalScaleComplete})
one gets
\begin{align}\label{OptimalScale}
    k_{\text{opt}} = \frac{-8R\xi_{G,1}G^{-2}_0-16\xi_{\tilde{\Lambda},1}-\left(g_0\xi_{g,1}\tilde{W}^2+4\tilde{B}^2g'_0\xi_{g',1}+8\xi_{m,1}-2B_{\mu}W^{\mu}_3(g'_0\xi_{g,1}+g_0\xi_{g',1})\right)\phi^2}{16RG_0^{-3}\zeta_1+32\xi_{\tilde{\Lambda},2}+\tilde{W}^2\phi^2\zeta_2+4\tilde{B}^2\phi^2\zeta_3+16\xi_{m,2}\phi^2-4B_{\mu}W^{\mu}_3\zeta_4\phi^2}.
\end{align}
This allows to write the optimal effective action~(\ref{Weakaction}) in a compact way, 
\begin{align}\label{OptimalEAwithCH}
    \Gamma_{\text{opt}}=\frac{1}{16}\int d^4x \sqrt{-g}\left\{32\tilde{\Lambda}_0-16\frac{R}{G_0}+16 (\partial_\mu \phi)(\partial^\mu \phi)+\left(16m^2_0+4\tilde{B}^2g'^2_0+g^2_0\tilde{W}^2-4B_{\mu}W^{\mu}_3 g_0 g'_0\right)\phi^2\nonumber
    \right. \\
    \left.
    -\frac{\left(8R\xi_{G,1}G^{-2}_0+16\xi_{\tilde{\Lambda},1}+8\xi_{m,1}\phi^2+g_0\tilde{W}^2\xi_{g,1}\phi^2+4\tilde{B}^2g'_0\xi_{h,1}\phi^2-2B_{\mu} W^{\mu}_3 (g_0\xi_{g',1}+g'_0\xi_{g,1})\phi^2\right)^2}{32\xi_{\tilde{\Lambda},2}16 R G^{-3}_0 (G_0 \xi_{G,2}-\xi^2_{G,1})+\tilde{W}^2\zeta_2\phi^2+4\tilde{B}^2\zeta_3\phi^2-4B_{\mu}W^{\mu}_3\zeta_4\phi^2}\right\},
\end{align}
where, 
\begin{subequations}
\begin{align}
    \zeta_1 &= G_0\xi_{G,2}-\xi^2_{G,1},\\
    \zeta_2 &= \xi^2_{g,1}+2g_0\xi_{g,2},\\
    \zeta_3 &= \xi^2_{g',1}+2h_0\xi_{g',2},\\
    \zeta_4 &= g'_0 \xi_{g,2}+\xi_{g,1} \xi_{g',2}+g_0 \xi_{g',2}.
\end{align}
\end{subequations}
In the weak field limit, one can expand the 
potential to order $\phi^4$. 
The Higgs parameters $\mu$ and $\lambda$ present in~(\ref{OptimalEA}) are now functions that depend only on the infrared scale $m_0$, the mass parameters $\xi_{m,1}, \xi_{m,2}$ and gravitational coupling parameters $\xi_{\tilde{\Lambda},1}, \xi_{\tilde{\Lambda},2}$,
\begin{subequations}\label{HiggsParameters}
\begin{align}
    \frac{\mu^2}{2} &= m^2_0 + \frac{\xi_{\tilde{\Lambda},1}(\xi_{\tilde{\Lambda},1}\xi_{m,2}-2\xi_{\tilde{\Lambda},2}\xi_{m,1})}{4\xi^2_{\tilde{\Lambda},2}}\\
    \frac{\lambda}{4} &= \frac{(\xi_{\tilde{\Lambda},1}\xi_{m,2}-\xi_{\tilde{\Lambda},2}\xi_{m,1})^2}{8\xi^3_{\tilde{\Lambda},2}}.
\end{align}
\end{subequations}
These two relations are analogous to the relations found in the 
 $\text{U}(1)$ Abelian case~\cite{Koch:2020baj}. 
 A requirement for the occurrence of spontaneous symmetry breaking in this model is $\xi_{\tilde{\Lambda},2}>0$. 
For SSB to occur, one needs $-\mu^2 > 0$ and $\lambda>0$.
In this case all the ground states all have
the same absolute value 
\begin{align}
    |\phi_0| \equiv \sqrt{\phi_0\phi_0^*} = \sqrt{\frac{-\mu^2}{2\lambda}}.
\end{align}
 Now, following our formalism, we can read-off from Eq.(\ref{HiggsParameters}b) that the requirement $\lambda>0$ implies $\xi_{\tilde{\Lambda,2}} > 0$ (given the that numerator in Eq.(\ref{HiggsParameters}b) is always possitive).

\subsection{Parameter expansion in the electroweak sector}

To reduce the number of unknown parameters, the $\xi$-values coming from the electroweak sector will be determined. 
The beta functions of the theory dictate the running of the couplings concerning the renormalization scale. As explained above, if the corresponding RG flow is analyzed in the infrared sector, one finds that the infrared divergences present in the electroweak sector do not allow an expansion around $k \rightarrow 0$ (unlike the case of gravity). Ignoring the gravitational contribution to the flow of the weak couplings and the anomalous dimension of the Higgs mass 
(such contributions are suppressed by inverse powers of the Planck mass $M^{-2}_{PL}$), the EW $\beta$-functions up to one loop in the minimal subtraction scheme~\cite{Bian:2013jra} read 
\begin{subequations}\label{BetaFunctions}
\begin{align}
    k\frac{\text{d}g}{\text{d}k} &= -\frac{19}{192\pi^2} g^3(k),\\
    k\frac{\text{d}g'}{\text{d}k} &= \frac{41 g'^3(k)}{96\pi^2},\\
    k\frac{\text{d}m_H}{\text{d}k} = -&\frac{{m_H(k)}}{16\pi^2}\left(\frac{3}{4}g^2+\frac{9}{4}g'^2\right).
\end{align}
\end{subequations}

These $\beta$-functions can be integrated between a low scale $k_0$ and an intermediate scale $k=k_0+ k'$.
In the vicinity of the IR scale
the running couplings $g_k, g'_k$ and $m_{H,k}$ are found to be
\begin{subequations}\label{CouplingExpansion2}
\begin{align}
    g(k) &= g_0-\frac{19g^3_0}{192\pi^2}\left(\frac{k'}{m_0}\right)+\frac{19(19g^5_0+64g^3_0\pi^2)}{24576\pi^4}\left(\frac{k'}{m_0}\right)^2+\mathcal{O}\left(\frac{k'}{m_0}\right)^3,\\
    g'(k) &= g'_0+\frac{41g'^3_0}{96\pi^2}\left(\frac{k'}{m_0}\right)+\frac{41(41g'^5_0-32g'^3\pi^2)}{6144\pi^4}\left(\frac{k'}{m_0}\right)^2+\mathcal{O}\left(\frac{k'}{m_0}\right)^3,\\
    m^2_H(k) &= m^2_0-\frac{3m^2_0(g^2_0+3g'^2_0)}{32\pi^2}\left(\frac{k'}{m_0}\right)+\frac{(28g^4_0+54g^2_0g'^2_0-165g'^4_0+96\pi^2(g^2_0+3g'^2_0))m^2_0}{2048\pi^4}\left(\frac{k'}{m_0}\right)^2+\mathcal{O}\left(\frac{k'}{m_0}\right)^3.
\end{align}
\end{subequations}

In the previous formulas, $g_0$ and $g'_0$ are the values of the EW couplings measured at the infrared scale. The set of expansion parameters from the EW sector thus can be identified from a comparison between (\ref{CouplingExpansion}) and (\ref{CouplingExpansion2}),

\begin{subequations}\label{ParametersExpansion}
\begin{align}
    \xi_{g,1}&=-\frac{19g^3_0}{192\pi^2},\\
    \xi_{g,2}&=\frac{19(19g^5_0+64g^3_0\pi^2)}{24576\pi^4},\\
    \xi_{g',1}&=\frac{41g'^3_0}{96\pi^2},\\
    \xi_{g',2}&=\frac{41(41g'^5_0-32g'^3\pi^2)}{6144\pi^4},\\
    \xi_{m,1}&=-\frac{3m^2_0(g^2_0+3g'^2_0)}{32\pi^2},\\
    \xi_{m,2}&=\frac{\left(28g^4_0+54g^2_0g'^2_0-165g'^4_0+96\pi^2(g^2_0+3g'^2_0)\right)m^2_0}{2048\pi^4}.
\end{align}
\end{subequations}

\subsection{Mass and vacuum expectation value of scalar field}

From the optimal action $\Gamma_{\text{opt}}$, one can read-off the terms proportional to $\phi^2$ and $\phi^4$. With this, the
resulting VEV and Higgs mass, utilizing  \eqref{HiggsParameters}, are: 
\begin{subequations}\label{Vev&Higgs1}
\begin{align}
    v^2 &= \frac{\xi_{\tilde{\Lambda},2}\left(\xi^2_{\tilde{\Lambda},1}\xi_{m,2}-2\xi_{\tilde{\Lambda},1}\xi_{\tilde{\Lambda},2}\xi_{m,1}+4 m^2_0 \xi^2_{\tilde{\Lambda},2}\right)}{(\xi_{\tilde{\Lambda},1}\xi_{m,2}-\xi_{\tilde{\Lambda},2}\xi_{m,1})^2},\\
    m^2_H &= \frac{\xi_{m,2}\xi^2_{\tilde{\Lambda},1}-2\xi_{m,1}\xi_{\tilde{\Lambda},1}\xi_{\tilde{\Lambda},2}+4m^2_0\xi^2_{\tilde{\Lambda},2}}{\xi^2_{\tilde{\Lambda},2}}.
\end{align}
\end{subequations}

Next, one can insert the set of $\xi$-values of the matter sector~(\ref{ParametersExpansion}) into (\ref{Vev&Higgs1}) to obtain,
\begin{subequations}\label{vevandhiggs}
\begin{align}
    v^2 &= \frac{2048 \pi^4 \xi_{\tilde{\Lambda},2}\left(28g_0^4\xi^2_{\tilde{\Lambda},1}-165g'^4_0\xi^2_{\tilde{\Lambda},1}+8192\pi^4\xi^2_{\tilde{\Lambda},2}+288g'^2_0\pi^2\xi_{\tilde{\Lambda},1}\mathbb{T}+6g^2_0\xi_{\tilde{\Lambda},1}(9g'^2_0\xi_{\tilde{\Lambda},1}+16\pi^2\mathbb{T})\right)}{m^2_0\left(28g^4_0\xi_{\tilde{\Lambda},1}+6g^2_0(9g'^2_0\xi_{\tilde{\Lambda},1}+16\pi^2\mathbb{S})+3g'^2_0(-55g'^2_0\xi_{\tilde{\Lambda},1}+96\pi^2\mathbb{S})\right)^2}\\
    m^2_H &= m^2_0\frac{28g_0^4\xi^2_{\tilde{\Lambda},1}-165g'^4_0\xi^2_{\tilde{\Lambda},1}+8192\pi^4\xi^2_{\tilde{\Lambda},2}+288g'^2_0\pi^2\xi_{\tilde{\Lambda},1}\mathbb{T}+6g^2_0\xi_{\tilde{\Lambda},1}(9g'^2_0\xi_{\tilde{\Lambda},1}+16\pi^2T)}{2048\pi^4\xi^2_{\tilde{\Lambda},2}},
\end{align}
\end{subequations}
where $\mathbb{T}= \xi_{\tilde{\Lambda},1}+4 \xi_{\tilde{\Lambda},2}$ and $\mathbb{S}= \xi_{\tilde{\Lambda},1}+2\xi_{\tilde{\Lambda},2}$.

\subsection{Mass of gauge fields}

The function $\mathcal{L}(W,B)$ includes (among others) the kinetic terms for the $\text{U(1)}_Y$ and $\text{SU(2)}$ gauge fields, necessary to extract the departure of $\Tilde{g}$ and $\Tilde{g'}$ from their infrared values $g_0$ and $g'_0$. The masses for the gauge fields are given by the lengthy expressions, 

\begin{align}\label{Wmass}
    m^2_W = \ &\frac{\xi^2_{\tilde{\Lambda},2}\mathcal{B}_3\left(4m^2_0+\xi^{-2}_{\tilde{\Lambda},2}\xi_{\tilde{\Lambda},1}(\xi_{m,2}\xi_{\tilde{\Lambda},1}-2\xi_{m,1}\xi_{\tilde{\Lambda},2})\right)}{256(\xi_{m,2}\xi_{\tilde{\Lambda},1}-\xi_{m,1}\xi_{\tilde{\Lambda},2})^4\mathcal{B}^2_1}   \times \nonumber\\
    & \left(-4g_0\xi_{g,1}\xi_{\tilde{\Lambda},2}\mathcal{B}_1\mathcal{B}_2+\xi^2_{g,1}\mathcal{B}^2_2+2g_0(2g_0\xi^2_{\tilde{\Lambda},2}\mathcal{B}^2_1+\xi_{g,2}\mathcal{B}^2_2)\right),
\end{align}
\begin{align}\label{Zmass}
    m^2_Z=\frac{(\xi_{m,2}\xi^2_{\tilde{\Lambda},1}-2\xi_{m,1}\xi_{\tilde{\Lambda},1}\xi_{\tilde{\Lambda},2}+4m^2_0\xi^2_{\tilde{\Lambda},2})^2}{256(\xi_{m,2}\xi_{\tilde{\Lambda},1}-\xi_{m,1}\xi_{\tilde{\Lambda},2})^4\mathcal{B}^2_1}\left\{4g^2_0\xi^2_{\tilde{\Lambda},2}\mathcal{B}^2_1-4\xi_{\tilde{\Lambda},2}\mathcal{B}_1\mathcal{B}_2\left(g_0\xi_{g,1}+4g'_0\xi_{g',1}\right) \xi_{\tilde{\Lambda},2}
    \right.\nonumber\\
    \left. +\left(\xi^2_{g,1}+2g_0\xi_{g,2}+4\xi^2_{g',1}\right)\mathcal{B}^2_2+8g'_0\left(2g'_0\xi^2_{\tilde{\Lambda},2}\mathcal{B}^2_1+\xi_{g',2}\mathcal{B}^2_2\right) \right\},
\end{align}
where the following definitions were used,
\begin{subequations}
    \begin{align}
        \mathcal{B}_1&=3\xi^2_{m,2}\xi^2_{\tilde{\Lambda},1}-6\xi_{m,1}\xi_{m,2}\xi_{\tilde{\Lambda},1}\xi_{\tilde{\Lambda},2}+2\xi^2_{\tilde{\Lambda},2}(\xi^2_{m,1}+2m^2_0\xi_{m,2}),\\
        \mathcal{B}_2&=2\xi^2_{m,2}\xi^3_{\tilde{\Lambda},1}-3\xi_{m,1}\xi_{m,2}\xi^2_{\tilde{\Lambda},1}\xi_{\tilde{\Lambda},2}+4m^2_0\xi_{m,1}\xi^3_{\tilde{\Lambda},2},\\
        \mathcal{B}_3&=\xi_{m,2}\xi^2_{\tilde{\Lambda},1}-2\xi_{m,1}\xi_{\tilde{\Lambda},1}\xi_{\tilde{\Lambda},2}+4m^2_0\xi^2_{\tilde{\Lambda},2}.
    \end{align}
\end{subequations}
Throughout this discussion, the mass associated with the photon remains 0. 
It is further necessary to note that the values of the masses $m_{H,W,Z}$ do not depend on some specific choice of the initial values for the gravitational couplings $G_0$, $\Lambda_0$ (the values of the gravitational couplings measured at the reference scale). This is an
advantage of our approach.
Remember that if physical masses would depend on these extreme gravitational couplings (e.g., the Planck mass $\sim 1/\sqrt{G_0}$), one would need a large fine-tuning to compensate for this effect and obtain moderate particle masses.

\subsection{Benchmark of gravitational parameters}

Once the masses of the scalar field and the gauge bosons are fixed, we 
can continue the analysis by exploring the possibility of selecting the weak couplings measured at the infrared scale $k_0$ to the known values $g_0=0.64$ and $g'_0=0.34$. 
Furthermore, the reference scale will be set using the measured mass of the Higgs boson. As a first approach, the value $m_H=125$~GeV 
will be considered as fix condition. Naturally, in a subsequent step, one can relax this condition to a mass range $m_H=(125\pm 0.14)$~GeV.
Imposing this condition, the reference mass scale as a function of the gravitational parameters is

\begin{align}\nonumber
m_H&=125~\text{GeV}\quad \Rightarrow\\ \label{m0fromHiggs1}
    m_0 &= \frac{2071.4\xi_{\Tilde{\Lambda},2}}{\sqrt{1.69\xi^2_{\Tilde{\Lambda},1}+6.71\xi_{\Tilde{\Lambda},1}\xi_{\Tilde{\Lambda},2}+1868.18\xi^2_{\Tilde{\Lambda},2}}}\text{GeV}.
\end{align}

\begin{figure}[!h]
\begin{subfigure}{.5\textwidth}
  \centering
  \includegraphics[width=.9\linewidth]{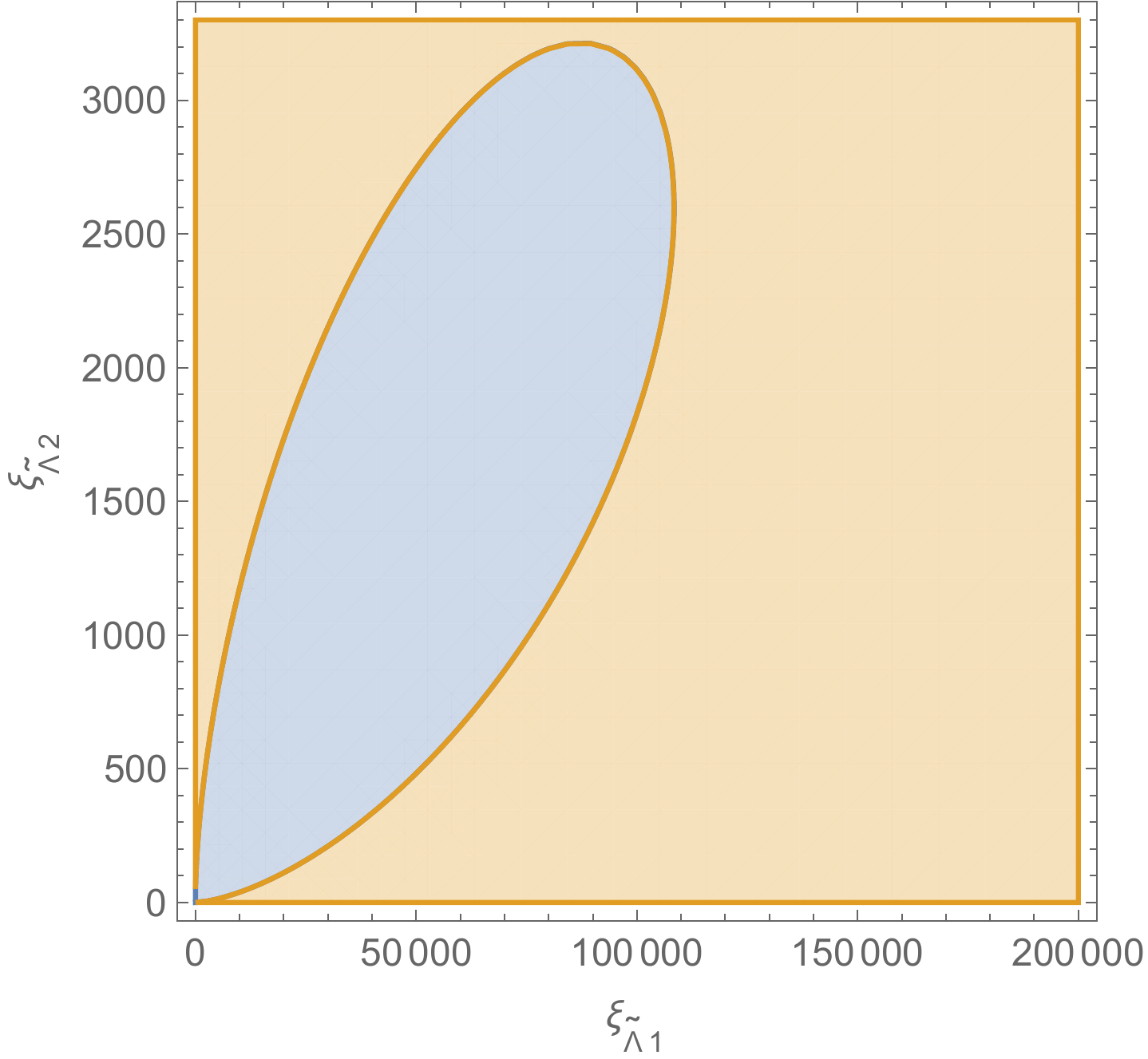}
  \caption{Region spanned by the W-boson masses~(\ref{Wmass})}
  \label{fig:sfig1}
\end{subfigure}%
\begin{subfigure}{.5\textwidth}
  \centering
  \includegraphics[width=.9\linewidth]{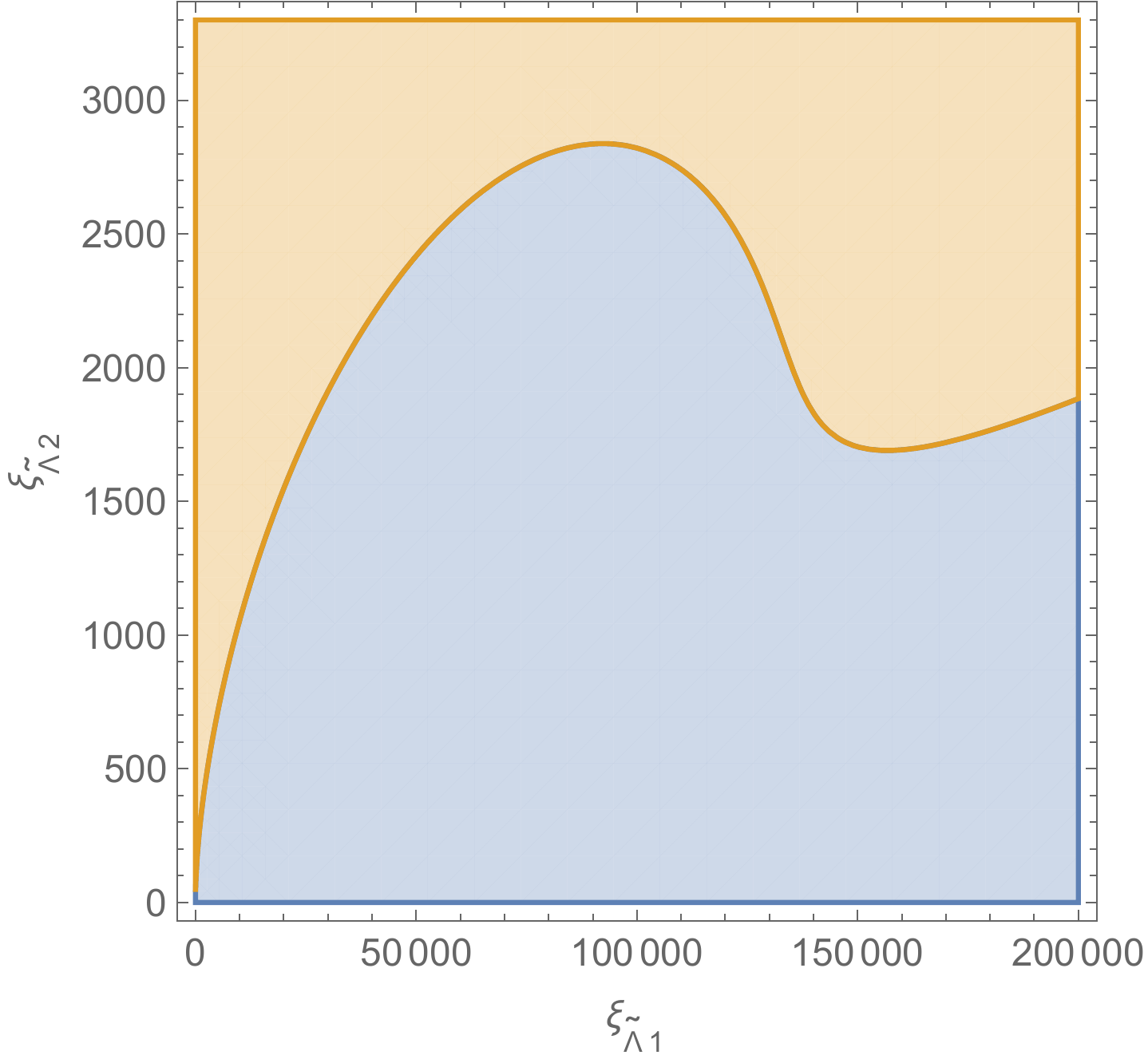}
  \caption{Region spanned by the Z-boson mass~(\ref{Zmass})}
  \label{fig:sfig2}
\end{subfigure}
\caption{Allowed parameter range in the gravitational parameter $\xi_{\tilde{\Lambda},1}$, $\xi_{\tilde{\Lambda},2}$.The orange region represents gravitational parameters in~(\ref{fig:sfig1}) 
which give a mass for the $SU(2)$ gauge bosons lower than $m_W=80.391$GeV, while the blue region represents parameters greater than $m_W=80.367$GeV. The same colors are used to limit the range for the $U(1)_Y$ gauge boson in~(\ref{fig:sfig2}), with the experimental upper ($m_Z=91.1897$GeV) and lower ($m_Z=91.1855$GeV) limits (see Table~\ref{Table1})}
\label{fig:fig}
\end{figure}

After replacing (\ref{m0fromHiggs1}) in the expressions for $m_W$~(\ref{Wmass}) and $m_Z$~(\ref{Zmass}), one  gets a system of two equations and two unknown gravitational parameters $\xi_{\tilde{\Lambda},1}$, $\xi_{\tilde{\Lambda},2}$. Since the observed masses for the gauge bosons have an experimental uncertainty, 
the solutions to these conditions are regions and not points in the parameter space.
The values where $\xi_{\tilde{\Lambda},1}$ and $\xi_{\tilde{\Lambda},2}$ meet the observational conditions, are given by  a finite region in the 
$\xi_{\tilde{\Lambda},1}\in\mathds{R}$, $\xi_{\tilde{\Lambda},2}>0$ parameter space. 
Figure~\ref{fig:sfig1} shows contours for the gravitational parameters $\xi_{\tilde{\Lambda},1},\xi_{\tilde{\Lambda},2}$ which are obtained by associating the limits of the experimental observed W-boson mass with the limits of $m_W$ in~(\ref{Wmass}).  
The exact process carried out with the Z-boson is shown in figure~\ref{fig:sfig2}.\\
\newline

\begin{table}[h]
    \begin{center}
        \begin{tabular}{|c|c|}
            \hline
            W boson mass ($m_W$) & $80.379 \pm\;0.012$ \text{ GeV} \\ \hline
            Z boson mass ($m_Z$) & $91.1876 \pm\;0.0021$ \text{ GeV} \\ \hline
            Higgs boson mass ($m_H$) & $125.18 \pm\;0.16$ \text{ GeV} \\ \hline
            Vacuum expectation value (VEV) & $246.220 \pm\;0.007$ \text{ GeV} \\ \hline
        \end{tabular}
    \end{center}
    \caption{Experimentally measured values for the bosons and Higgs vacuum expectation value~\cite{ParticleDataGroup:2018ovx}}\label{Table1}
\end{table}

\newpage 

\begin{figure}[!h]
  \includegraphics[width=1\linewidth]{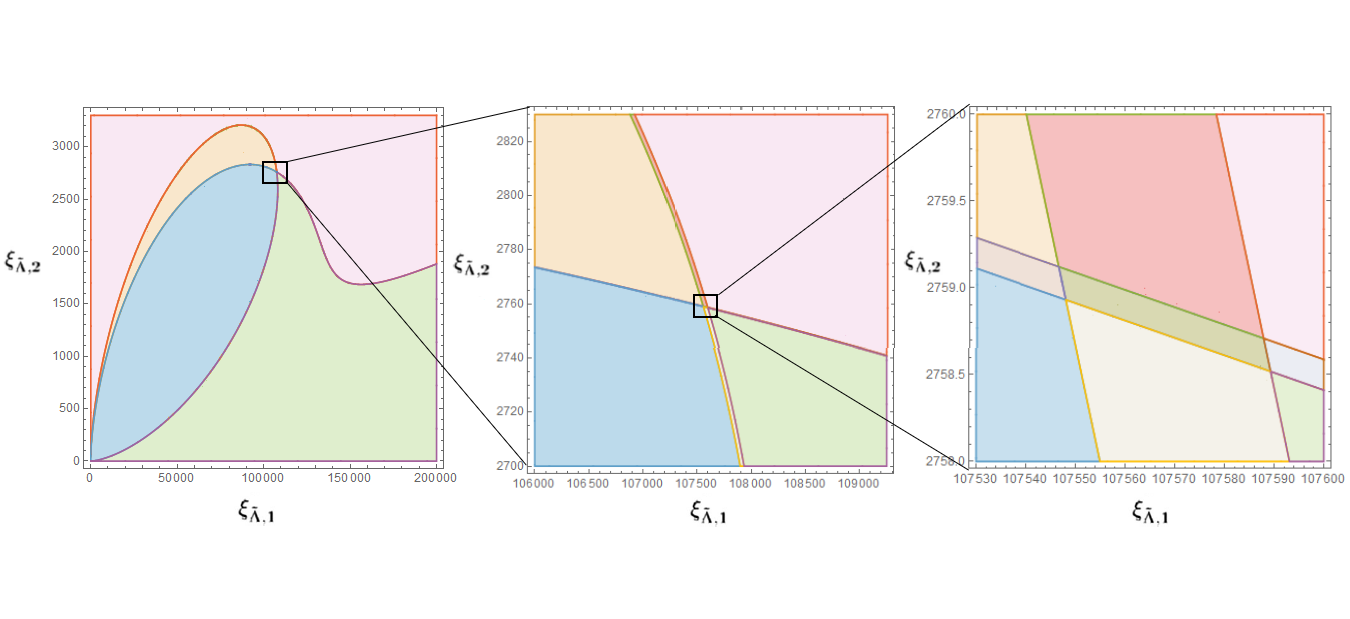}
  \caption{Overlapping and magnification on Figure~\ref{fig:sfig1} and Figure~\ref{fig:sfig2}. The gravitational parameters in the intersection dark green region allow the experimental results for the observed masses of the Higgs, W, and Z bosons. Table~\ref{Tab.label1} contains the color-coding
  for the experimental requirements of this figure.}
  \label{fig:Intersection}
\end{figure}

The overlap between the two bands of the allowed gravitational parameters leads to a zone of gravitational parameters that fulfill all given experimental constraints on the scalar and gauge boson masses. This outcome is displayed in figure~\ref{fig:Intersection}.

\begin{figure}[h]
  \centering
  \includegraphics[width=0.8\linewidth]{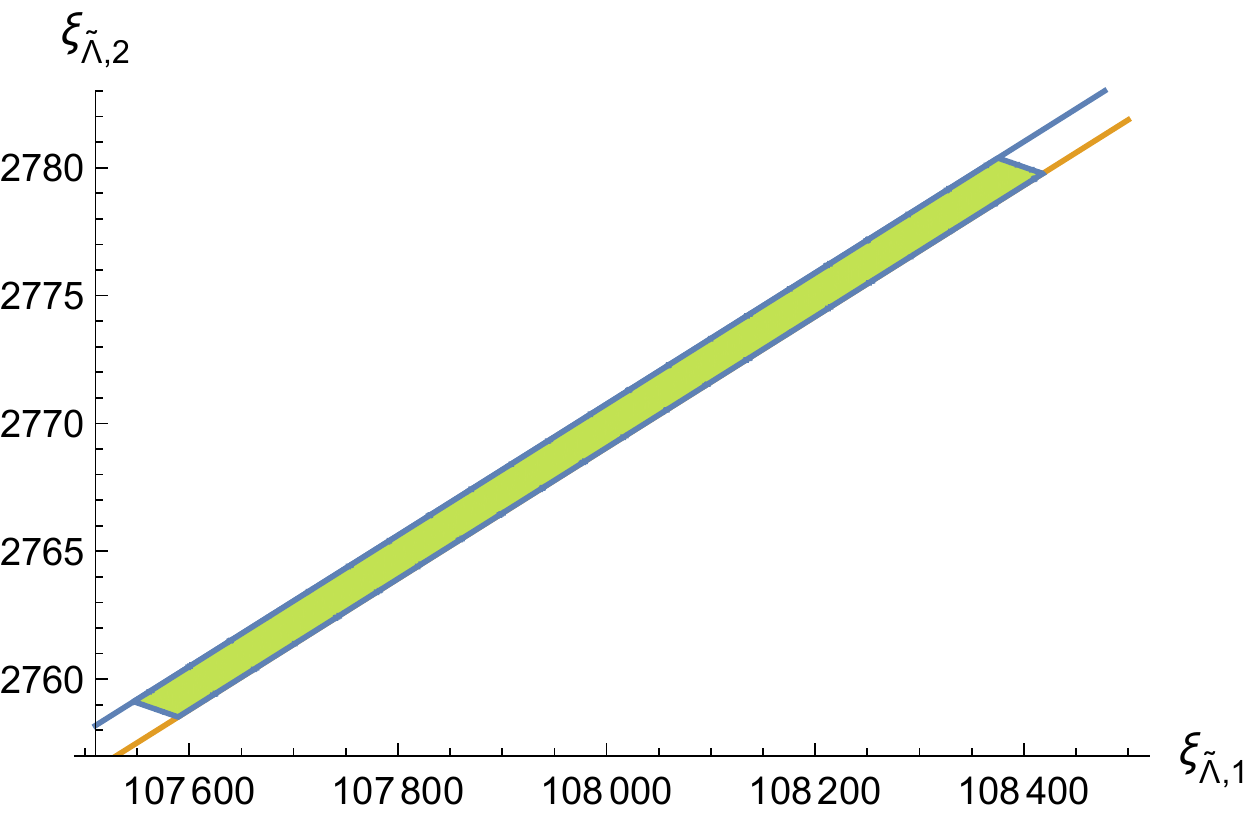}
\caption{Section in the parameter space $\xi_{\tilde{\Lambda},1},\xi_{\tilde{\Lambda},2}$ where the spontaneous symmetry breaking 
generated by 
meets all the experimental requirements imposed on the observed values $m_H, m_W$ and $m_Z$. 
The plot was obtained after considering all the experimental permitted values of the Higgs boson masses in~(\ref{m0fromHiggs1}).}
\label{regions2}
\end{figure}

As one varies the values of the Higgs mass used to set the infrared scale $m_0$ in~(\ref{m0fromHiggs1}) through the experimentally observed values, the intersection zone encountered in figure~\ref{fig:Intersection} is slightly shifted in the space spanned by the gravitational parameters. 
This leads to an increased intersection area, as shown in figure~\ref{regions2}.
\begin{table}[t!]
	\centering
	\begin{tabular}{|c||c|c|}
	\hline
	 & $m_W[\text{GeV}]$ & $m_Z[\text{GeV}]$ \\
	\hline
	\centering\fcolorbox{black}{OliveDrab3} & $m^-_W \leq m_W \leq m^+_W$ & $m^-_Z \leq m_Z \leq m^+_Z$ \\ \hline
	\fcolorbox{black}{LightGoldenrod2} & $m_W<m^-_W$ & $m_Z>m^+_Z$ \\ \hline
	\fcolorbox{black}{Tomato1} & $m^-_W \leq m_W \leq m^+_W$ & $m_Z>m^+_Z$ \\ \hline
	\fcolorbox{black}{MistyRose1} & $m^+_W < m_W$ & $m^+_Z < m_Z$ \\ \hline
	\fcolorbox{black}{SkyBlue1} & $m_W<m^-_W$ & $m_Z<m^-_Z$ \\ \hline
	\fcolorbox{black}{AntiqueWhite2} & $m_W<m^-_W$ & $m^-_Z \leq m_Z \leq m^+_Z$ \\ \hline
	\fcolorbox{black}{LightCyan1} & $m^+_W<m_W$ & $m^-_Z \leq m_0 \leq m^+_Z$ \\ \hline
	\fcolorbox{black}{LemonChiffon1} & $m^-_W \leq m_W \leq m^+_W$ & $m^+_W<m_W$ \\ \hline
	\fcolorbox{black}{DarkSeaGreen1} & $m_Z<m^-_Z$ & $m_Z<m^-_Z$ \\ \hline
	\end{tabular}
	\caption{\label{Tab.label1} 
	Color-coding
	for the nine regions that characterize figure~\ref{regions2}. 
	The experimental data used in the benchmark is taken from table 1, where the quantities $m^+_W=80.367$ GeV, $m^-_W=80.391$ GeV, $m^-_Z=91.1855$GeV and $m^+_Z=91.1897$GeV were defined as the minimum and maximum values that the experimental values of the gauge boson masses can take.}
\end{table}

Only the tiny fraction of parameters shown in figure \ref{regions2} produces the SSB with the observed masses and couplings. This region of gravitational RG parameters is thus phenomenologically preferred. It is fascinating to check whether this is a viable region from the FRG point of view. This question will be addressed in the following section.
%

\section{Comparison with the Functional Renormalization Group}

The SSB mechanism obtained in the previous sections employs the vacuum energy density as a central object. 
The cosmological constant term allows a well-defined effective symmetry breaking potential even in flat spacetime, providing a constant term independent of the scalar field.

In the infrared expansion~(\ref{CouplingExpansion}), 
the gravitational constants expansion is left arbitrary, so we are leaving unknown the energy-dependence of the gravitational running couplings. 

In a phenomenological approach, of the infrared expansion~(\ref{CouplingExpansion}) the gravitational constants  ($\xi_{\tilde{\Lambda},1}$ y $\xi_{\tilde{\Lambda},2}$) can be left 
considered as free parameters.
As a result, the VEV and gauge boson masses obtained in~(\ref{vevandhiggs})-(\ref{Zmass}) do not rely on a particular election of the quantum gravity scenario, even when they strongly depend on the infrared expansion of $\tilde{\Lambda}_k$. 
Thus, the corresponding expressions are written in terms of the parameters $\xi$s. 
The formalism used here implies that the coupling constants $\{ G_k, \tilde{\Lambda}_k, m^2_k, g'_k, g_k \}$ can be expanded in the vicinity of a particular reference scale $m_0$. This feature is quite common in a variety of scenarios in QG. In such a sense, it should be reinforced that we are not taken a concrete scenario of QG. Albeit possible, our formalism and idea try to face the problem irrespectively of the concrete QG theory.

In the following, the condition imposed on $\tilde{\Lambda}_k$ ($\xi_{\tilde{\Lambda},2}>0$) will be tested with the results obtained by the Functional Renormalization Group (FRG) approach to quantum gravity, the so-called Asymptotic Safety (AS) scenario.

The evolution of the scale-dependent couplings is dictated by the FRG equation~\cite{Wetterich:1992yh,Morris:1993qb},
\begin{equation}\label{eq:Wetterich}
    \frac{\text{d}\Gamma_{k}}{\text{d}t}=\frac{1}{2}\text{STr}\left(\frac{\partial_{t}\mathcal{R}_{k}}{\Gamma^{(2)}_{k}[\phi]+\mathcal{R}_{k}}\right).
\end{equation}
where $t$ denotes the renormalization group time $t \equiv \ln(k/k_0)$ and $k_0$ represents an arbitrary reference scale. The supertrace STr represents a sum over all internal indices as well as an integration over spacetime. The argument of the trace involves the regulator function $\mathcal{R}_k$, an arbitrary function of the eigenvalues of the Bochner Laplacian that falls off sufficiently fast for $p^2>k^2$. 
For $p^2<k^2$, $\mathcal{R}_k$ acts as a mass term, suppressing these modes and implementing the idea of performing the functional integral one momentum shell at a time. Finally, the modified inverse propagator requires the evaluation of the second functional derivative of $\Gamma_k$ concerning the fluctuation fields, which typically is matrix-valued in field space.

To perform the comparison between~(\ref{vevandhiggs}-\ref{Zmass}) and the FRG~(\ref{eq:Wetterich}), the pure gravity scenario~\cite{Reuter:1996cp,Reuter:2001ag}
must be supplemented by matter degrees of freedom. The compatibility of the Reuter fixed point with minimally and nonminimally coupled matter fields has been investigated within AS for a long time \cite{Griguolo:1995db,Percacci:2002ie,Zanusso:2009bs,Percacci:2003jz,Narain:2009fy,Narain:2009gb,Shaposhnikov:2009pv,Folkerts:2011jz,Eichhorn:2011pc,Dona:2013qba,Dona:2014pla,Labus:2015ska,Meibohm:2015twa,Eichhorn:2016esv,Hamada:2017rvn,Biemans:2017zca,Eichhorn:2017sok,Alkofer:2018fxj,Alkofer:2018baq,Pawlowski:2018ixd,Wetterich:2019rsn,Daas:2020dyo,Daas:2021abx,Laporte:2021kyp,Ohta:2021bkc}. In this section, the truncation is taken from~\cite{Dona:2013qba,Dona:2014pla,Dona:2015tnf,Eichhorn:2017egq}. The matter content consists of $N_{s}$ scalar fields $\phi^{i}$, $N_{D}$ fermion fields $\psi^{i}$, and $N_{V}$ Abelian gauge fields $A^{i}_{\mu}$. In addition to this, one has
the ghost and antighost fields $c$ and $\bar{c}$, all coupled to an external metric $g_{\mu\nu}$. The matter part of the action is given by,
\begin{eqnarray}
\label{GammaMatter}\Gamma_{matter}&=&S_{S}(\phi,g)+S_{D}(\psi,\bar{\psi},g)+S_{V}(A,c,\bar{c},g),\\
S_{S}(\phi,g)&=&\frac{1}{2}\int d^dx\sqrt{g}g^{\mu\nu}\sum^{N_{S}}_{i}\partial_{\mu}\phi^{i}\partial_{\nu}\phi^{i},\\
S_{D}(A,c,\bar{c},g)&=&i\int d^dx\sqrt{g}\sum^{N_{D}}_{i}\bar{\psi}_{i}{D\!\!\!\!/}\psi^{i},\\
S_{V}(A,c,\bar{c},g)&=&\frac{1}{4}\int d^dx \sqrt{g}\sum^{N_{V}}_{i=1}g^{\mu\nu}g^{\kappa\lambda}F^{i}_{\mu\kappa}F^{i}_{\nu\lambda}
+\frac{1}{2\xi}\int d^dx \sqrt{g}\sum^{N_{V}}_{i=1}\left(g^{\mu\nu}\nabla_{\mu}A^{i}_{\nu}\right)^2,\nonumber\\
&&+\int d^dx \sqrt{g}\sum^{N_{V}}_{i=1}\bar{c}_{i}(-\nabla^2)c_{i},
\end{eqnarray}
where ${D\!\!\!\!/}=\gamma^{a}e^{\mu}_{a}\nabla_{\mu}$, with the orthonormal frame $e^{\mu}_{a}$ and where \textit{i} is a summation index over matter species. The gravitational beta functions that dictate the evolution of the dimensionless couplings with respect to the renormalization scale $k$ are, 
\begin{subequations}\label{BetaFunctions2}
\begin{align}
        \label{eq:twentyfiveone}
        \beta_{g}=&2g+\frac{g^2}{6\pi}\left(N_{S}+2N_{D}-4N_{V}-46\right),
        \\
        \label{eq:twentyfivetwo}
        \beta_{\lambda}=&-2\lambda+\frac{g}{4\pi}(N_{S}-4N_{D}+2N_{V}+2) \nonumber\\ 
        &+\frac{g\lambda}{6\pi}\left(N_{S}+2N_{D}-4N_{V}-16\right).
\end{align}
\end{subequations}
where $\beta_g$ and $\beta_\lambda$ were expanded up to second order in $g$ and $\lambda$. For comparing the sign of $\xi_{\tilde{\Lambda},2}$ with~(\ref{BetaFunctions2}), the expansion of the dimensionful gravitational couplings is needed. It is convenient to define the following notation, 
\begin{subequations}\label{eq:AuxiliaryConstants}
    \begin{align}
            \label{eq:twentyeight.one}
            \mathcal{C}_{1}=& \frac{1}{6\pi}\left(N_{S}+2N_{D}-4N_{V}-46\right),
            \\
            \label{eq:twentyeight.two}
            \mathcal{C}_{2}=& \frac{1}{4\pi}\left(N_{S}-4N_{D}+2N_{V}+2\right),
            \\
            \mathcal{C}_{3}=&\frac{1}{6\pi}\left(N_{S}+2N_{D}-4N_{V}-16\right).
    \end{align}
\end{subequations}
The set of equations~(\ref{BetaFunctions}) can be integrated analytically. Starting with~(\ref{eq:twentyfiveone}), the dimensionful running Newton's coupling turns out to be,
\begin{equation}
    \frac{G_k}{2+\mathcal{C}_1 \, k^2 \, G_k} = \mathbb{C}_g,
\end{equation}
where $\mathbb{C}_g$ is an integration constant. If one is interested in the deep infrared regime $k\rightarrow 0$, the integration constant will be $\frac{G_N}{2}$, with $G_N=G(k\rightarrow 0)$. However, since the running couplings in this work were expanded at the infrared scale $m_0$~(\ref{CouplingExpansion}), the same procedure will be carried out to obtain a reliable comparison. Denoting $G_0$ the Newton coupling measured at $m_0$, one gets
\begin{equation}
    \frac{G_k}{2+\mathcal{C}_1 \, k^2 \, G_k} = \frac{G_0}{2+\mathcal{C}_1 \, m^2_0 \, G_0}.
\end{equation}
The differential equation~(\ref{eq:twentyfivetwo}) also admits an analytical integration, 
\begin{equation}
    \frac{G_k \, \Lambda_k}{(2 + \mathcal{C}_1 \, k^2 \, G_k)^{1+\frac{\mathcal{C}_3}{\mathcal{C}_1}}} = -\frac{\mathcal{C}_2}{\mathcal{C}_3 (\mathcal{C}_1+\mathcal{C}_3)} \, \frac{2 + k^2 \, G_k \, (\mathcal{C}_1+\mathcal{C}_3)}{(2 + \mathcal{C}_1 \, k^2 \, G_k)^{1 + \mathcal{C}_1 + \frac{\mathcal{C}_3}{\mathcal{C}_1}}} + \mathbb{C}_\lambda.
\end{equation}
with $\mathbb{C}_\lambda$ an integration constant. After the evaluation of $\mathbb{C}_\lambda$ at $k \rightarrow m_0$, $\Lambda_k$ is given by,
\begin{align}
\begin{split}
    \Lambda_k =&\left(\frac{G_0 \, \Lambda_0}{G_k} + \frac{\mathcal{C}_2}{\mathcal{C}_3 (\mathcal{C}_1 + \mathcal{C}_3)} \left(\frac{2 + m^2_0 \, G_0 \, (\mathcal{C}_1 + \mathcal{C}_3)}{G_k}\right)\right) \left(\frac{2 + \mathcal{C}_1 \, k^2 \, G_k}{2 + \mathcal{C}_1 \, m^2_0 \, G_0}\right)^{1 + \frac{\mathcal{C}_3}{\mathcal{C}_1}}\nonumber\\
    &+\frac{\mathcal{C}_2}{\mathcal{C}_3 (\mathcal{C}_1 + \mathcal{C}_3)} \left(\frac{2 + k^2 \, G_k (\mathcal{C}_1+\mathcal{C}_3)}{G_k}\right).
\end{split}
\end{align}
The expansion of the modified running $\tilde{\Lambda}_k$ around the quantity $k=(k'+m_0)/m_0$, with $k' \rightarrow 0$ up to order $k'^2$ reads, 
\begin{equation}
    \tilde{\Lambda}_k = \frac{\Lambda_0}{G_0} + \mathbb{E}_1 \, \left(\frac{k'}{m_0}\right) + \mathbb{E}_2 \, \left(\frac{k'}{m_0}\right)^2 + \mathcal{O}\left(\frac{k'}{m_0}\right)^3,
\end{equation}
where the constants $\mathbb{E}_1$ and $\mathbb{E}_2$ are defined as, 
\begin{subequations}\label{EConstants}
    \begin{align}
            \label{E1}
            \mathbb{E}_{1}=& m^2_0\left(\mathcal{C}_2 \, m^2_0 - \mathcal{C}_1 \, \Lambda_0 + \mathcal{C}_3 \, \Lambda_0\right),\\
            \label{E2}
            \mathbb{E}_{2}=& \frac{m^2_0}{2} \left[3 m^2_0 \, \mathcal{C}_2 + \left(\mathcal{C}_3-\mathcal{C}_1\right) \, \left(\Lambda_0 + G_0 \, m^4_0 \, \mathcal{C}_2 + G_0 \, \Lambda_0 \, m^2_0 \, \mathcal{C}_3\right)\right].
    \end{align}
\end{subequations}
Note that in the deep infrared regime ($m_0\rightarrow 0$), $\tilde{\Lambda}$ reduces to $\frac{\Lambda_I}{G_N}$ plus corrections of order $k'^2$. The comparison between~(\ref{E2}) and~(\ref{GammaTildeExpanssion}) reveals that the stability of the symmetry breaking potential implies a positive defined $\mathbb{E}_2$-value. Irrespective of the (real) value of the infrared scale, the first term on the right-hand side of~(\ref{E2}) dominates, so the condition $\xi_{\tilde{\Lambda},2}>0$ implies, 
\begin{equation}\label{RequirementModel}
    \mathcal{C}_2 > 0.
\end{equation}
To get the full comparison with the model~(\ref{GammaMatter}), the condition~(\ref{RequirementModel}) should be contrasted with two additional requirements needed in AS, namely,
\begin{enumerate}
    \item \textit{Positive Newton fixed point $g^*>0$}.-The low value ($k\lesssim
M_{pl}$,) of Newtons' gravitational coupling, is restricted by observations based on laboratory experiments at the scale $k_{lab}\backsimeq
 10^{-5} eV$~\cite{Gubitosi:2018gsl}.
    \item \textit{Two relevant directions}.-Insofar as the corresponding fixed points for the gravitational couplings of a pure gravity theory have two relevant directions \cite{Dona:2013qba}, one expects that the addition of a small number of matter degrees of freedom does not change this behavior and the subsequent parametrization in theory space.
    For additional details regarding how fixed points and critical exponents evolve, please consult Table \eqref{ta:morse}. 
\end{enumerate}
The previous points and the condition~(\ref{RequirementModel}) impose restrictions on the field content such that the proposed mechanism agrees with some of the requirements coming from AS.
We have collected in Table \eqref{ta:morse} specific fixed points and their relevant directions for different values of the set $\{N_s, N_D\}$ assuming $N_V=4$. At this point, we can notice a few features:
i) for $Ns = 4 = N_V$, for an increasing value of $N_D$ the dimensionless Newton's coupling increases and the dimensionless cosmological coupling decreases until it reaches negative values.

\begin{figure}[h!]
  \centering
  \includegraphics[width=0.7\linewidth]{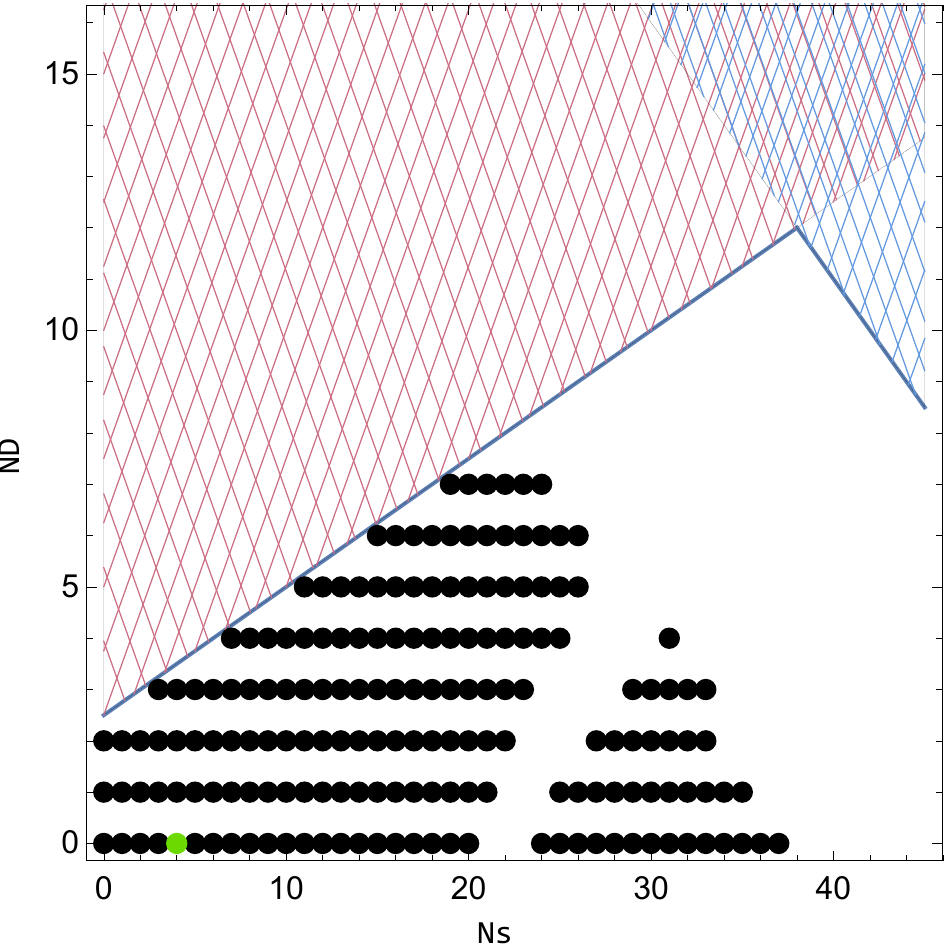}
\caption{Matter content allowed by the condition~(\ref{RequirementModel}), two relevant directions, and a positive definite gravitational Newton's fixed point. The red zone contains points with $g*<0$, while the shaded blue and yellow regions have $\xi_{\tilde{\Lambda},2}<0$. Black bullets represent the allowed dynamical matter degrees of freedom with two relevant directions. The green bullet represents the matter content of the model analyzed in the previous section.}
\label{ComparisonAS}
\end{figure}

Also, the first critical exponent, $\theta_1$, tends to increase with $N_D$ in most regions of parameter space.
Respect to the second critical exponent, $\theta_2$, a quite different behaviour is observed, i.e., for small values of $N_D$, 
it starts from its higher value and, when $N_D$ increases, $\theta_2$ decreases. 
For $N_D = 4$, its value increases and, subsequently, $\theta_2$ continues decreasing.
ii) for $N_V=4$, $N_D$ fixed, 
the dimensionless Newton's coupling and the dimensionless cosmological coupling increase with $N_s$. 
However, it is essential to mention that both of them could have a point at which this tendency is interrupted (see, for instance, $N_s=20$, $N_D=0$, where the dimensionless coupling constant are $\{ g^*, \lambda^* \} = \{ 0 , 0 \}$).
Now, concerning the behavior of the critical exponents, one should notice that $\theta_1$, in general, increases with $N_s$. However, as can be observed in Table \eqref{ta:morse}, one always has a point where $\theta_1$ is lower than its previous value. 
The second critical exponent shows more erratic behavior. Notably, one observes that depending on the value of $N_s$ (for a given $N_D$ and $N_V$), the value of $\theta_2$ could increase or decrease.
These conditions are pictured out in figure~\ref{ComparisonAS}, for $N_V=4$. For a fixed number of scalar degrees of freedom ($N_S=4$ for the standard model case), the inclusion of sterile fermions could destroy the requirement $\lambda>0$, then generating an unbounded potential from below. 
The inclusion of interacting fermionic degrees of freedom and gauge bosons responsible for the strong force between quarks will be left to future work. 
%
One can go one step further and integrate the beta functions up to $m_0$ to see if initial conditions exist for the RG trajectories intersecting the region spanned in figure~\ref{regions2}. 
The flow of $g_k$ and $\lambda_k$ can be integrated analytically to obtain the functional dependence of the couplings with the scale $k$. 
Current measurements of the gravitational Newton constant are based on laboratory experiments performed at energy scales in the range of $10^{-4} - 10^{-6} \, \text{GeV}$, giving $G^{\text{lab}}=6.7 \, {\text{x}} \, 10^{-57}$~\cite{ParticleDataGroup:2016lqr}. 
For the cosmological constant, the measurements are performed at energy scales of order $k_{\text{hub}} \approx 10^{-33} \, \text{GeV}$~\cite{Planck:2018vyg}. Based on the standard $\Lambda$CDM cosmological model, 
the estimation of the cosmological constant is of the order 
$\Lambda \approx 4 \times 10^{-66}$ 
eV$^2$.
%
In the present case, the infrared scale has not been set to a fixed value, but the evolution of $m_0$ depends on the VEV of the scalar field. Furthermore, the initial conditions of the gravitational couplings are evaluated at $m_0$ instead of $k_{\text{lab}}$ or $k_{\text{hub}}$; therefore we expect different numerical values for $G_0 = G(k\rightarrow m_0)$ and $\Lambda_0 = \Lambda(k \rightarrow m_0)$. Given the freedom in $m_0$ and the values of the dimensionful gravitational couplings at this scale, one can scan for which values (if any) the resulting curves intersect with the region found in figure~\ref{regions2}. 
From figure~\ref{regions2} and table~\ref{Tab.label2}, we observe that such values exist with and without the need to include sterile fermions in the field content of~(\ref{BetaFunctions}). This fact is highly non-trivial.
Table~\ref{Tab.label2} indicates that, given such an intersection, the value of the corresponding $m_0$ is correlated 
of the corresponding values of $G_0$ and $\Lambda_0$.
Small values of $m_0$ come at the price of large values of $G_0, \Lambda_0$, and vice versa.
\begin{figure}[ht!]
  \centering
  \includegraphics[width=.49\linewidth]{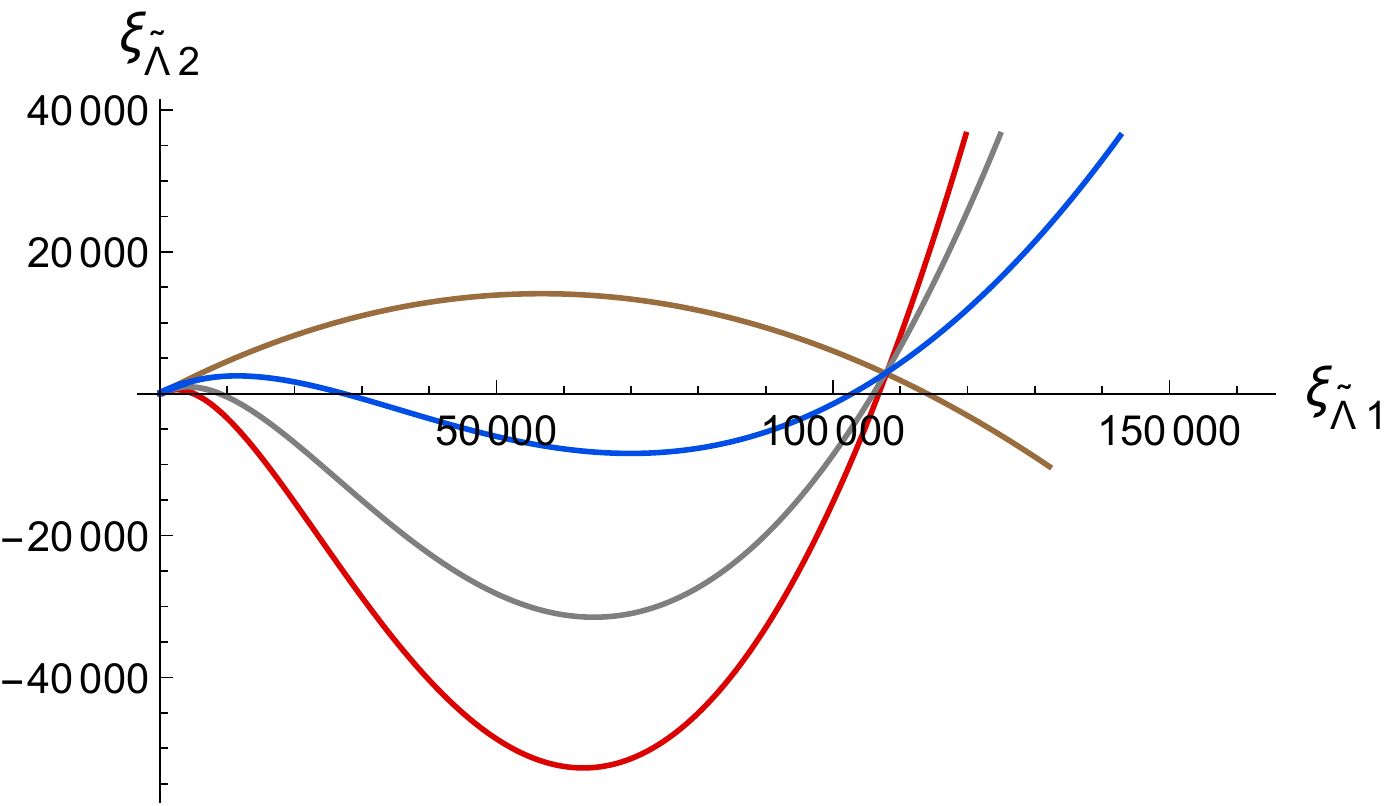}
  \includegraphics[width=.49\linewidth]{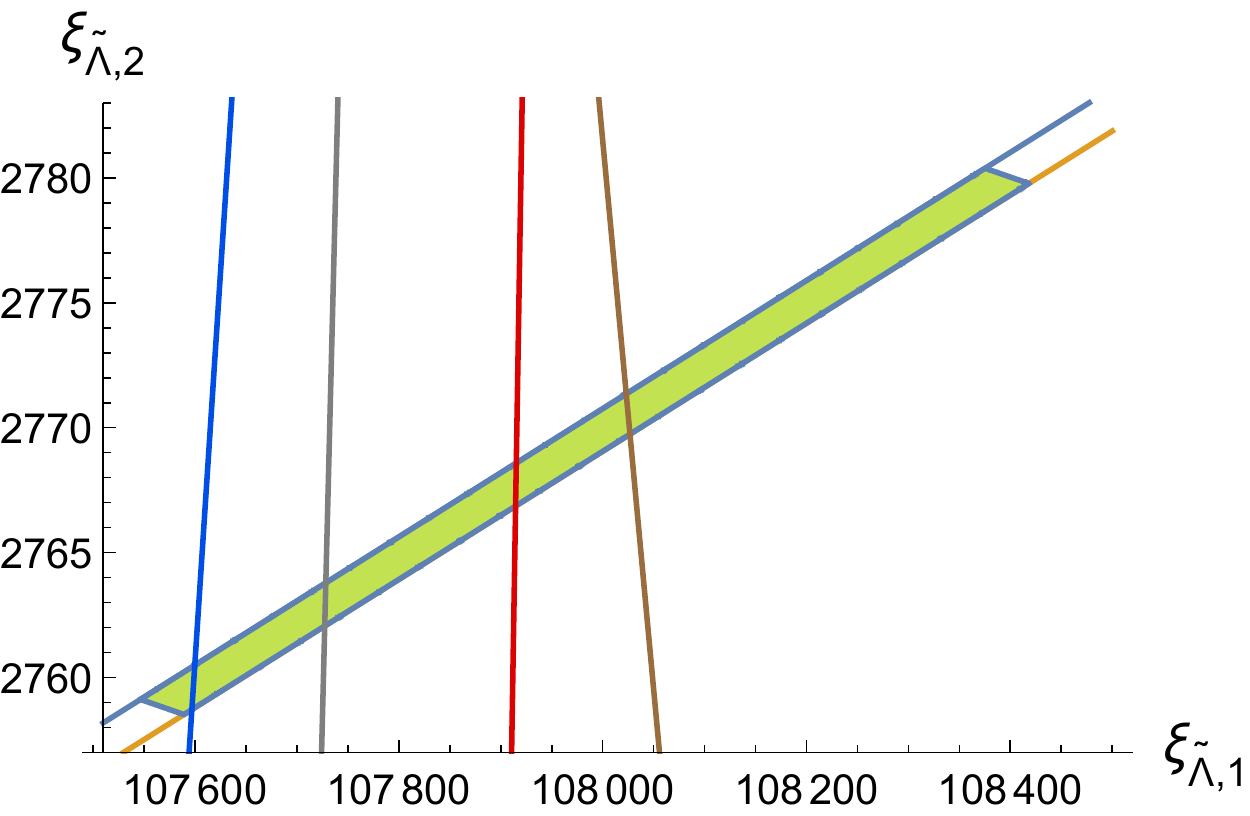}\\
    \includegraphics[width=.49\linewidth]{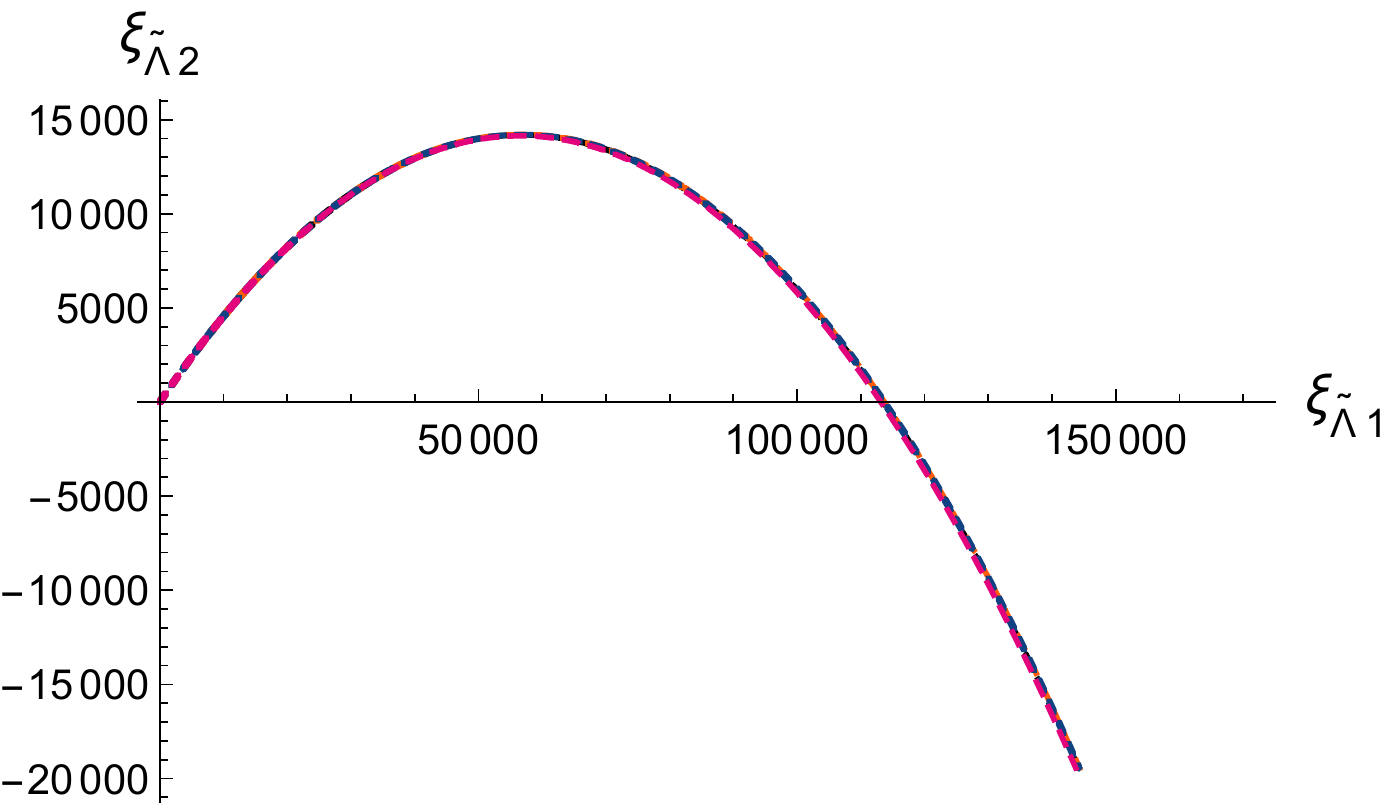}
  \includegraphics[width=.49\linewidth]{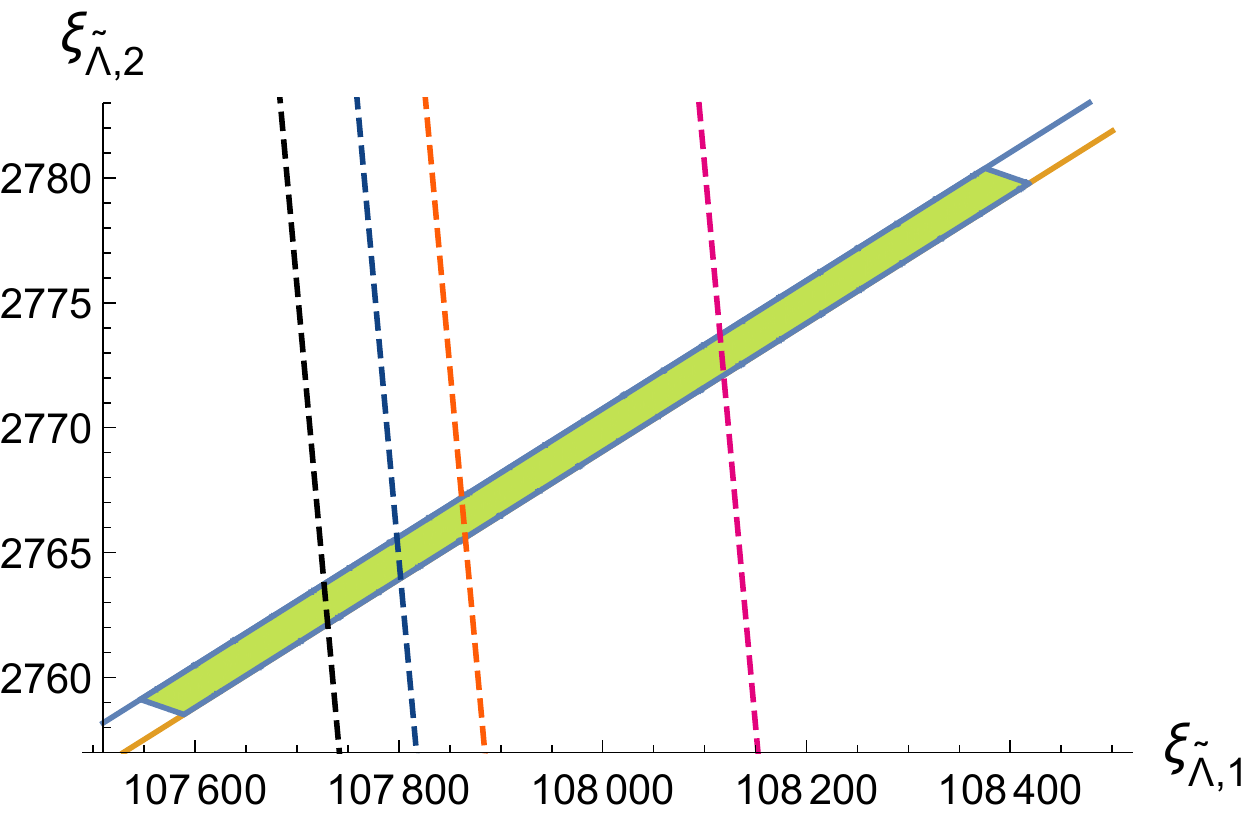}
  \caption{RG trajectories of the EH truncation with $\Gamma_{\text{matter}}$ minimally coupled to gravity at one-loop approximation (without anomalous dimensions for different species). Each curve corresponds to different values of $m_0$, $G_0$, $\Lambda_0$, and sterile fermions (see Table 3). The left panel shows the values associated with different initial conditions for several numbers of fermions that respect the bounds imposed by~\ref{ComparisonAS}, while the figures in the right panel despite the intersection of these curves of the parameter space~\ref{regions2}}
\label{RGfromAS}
\end{figure}
\begin{table}[t!]\label{Table3}
	\centering
	\begin{tabular}{c||c|c|c|c|}
	\hline
	 & $m_0[\text{GeV}]$ & $G_0[\text{GeV}]$ & $\Lambda_0[\text{GeV}]$ & $N_D$ \\
	\hline
	\centering\fcolorbox{black}{Burlywood3} & $0< m_0 \leq 5.5$ & $0.0265$ & $2723.6094$ & $0$ \\ \hline
	\fcolorbox{black}{Firebrick2} & $0< m_0 \leq 17$ & $0.052$ & $179.7188$ & $0.5$ \\ \hline
	\fcolorbox{black}{Azure3} & $0< m_0 \leq 17$ & $0.0323$ & $163.9688$ & $1.5$ \\ \hline
	\fcolorbox{black}{DodgerBlue1} & $1\leq m_0 \leq 20$ & $0.0139$ & $144.125$ & $2.5$ \\ \hline
	\fcolorbox{black}{black} & $1\leq m_0 \leq 100$ & $1.289\text{x}10^{-4}$ & $9.9815$ & $3.5$ \\ \hline
	\fcolorbox{black}{SteelBlue3} & $1\leq m_0 \leq 600$ & $3.764\,\text{x}10^{-6}$ & $0.2994$ & $3.5$ \\ \hline
	\fcolorbox{black}{Orange1} & $1\leq m_0 \leq 1150$ & $8.779 \, \text{x} \, 10^{-7}$ & $6.987 \, \text{x} \, 10^{-2}$ & $3.5$ \\ \hline
	\fcolorbox{black}{VioletRed2} & $1\leq m_0 \leq 18000$ & $3.766 \, \text{x} \, 10^{-9}$ & $3.004 \, \text{x} \, 10^{-4}$ & $3.5$ \\ \hline
	\end{tabular}
	\caption{\label{Tab.label2} Color coding  of Figure~\ref{RGfromAS} for different values of $m_0$, sterile fermions, and the gravitational parameters $G_0$, $\Lambda_0$. The gravitational parameters are left arbitrary since they are evaluated at an intermediate scale $m_0$ and not the deep infrared $k \rightarrow 0$}
\end{table}
\pagebreak

\section{Discussion and Conclusion}
\subsection{Summary}

In a recently reported paper, we have proposed a novel mechanism that has the potential to generate spontaneous symmetry breaking,
even in the absence of a tree-level potential for the scalar field~\cite{Koch:2020baj}.
The key ingredients of this novel approach are the interplay between quantum (gravitational) scale-dependence and 
the method of variational scale setting.
While in~\cite{Koch:2020baj} we tested this mechanism for a QED toy model, in the present paper, the study
a much more complete electroweak sector with a scalar doublet, but without the usual ad-hoc potential for this doublet.

For this purpose, we parametrize the IR-tale of the quantum (both electroweak and gravitational) coupling flow~(\ref{CouplingExpansion}).
Based on this parametrization, we perform the scale-setting, and weak field expansion proposed in~\cite{Koch:2020baj}.
By doing this, we can notice several non-trivial features discussed in deep throughout the manuscript. One of the most remarkable characteristics is the appearance of an (effective) Higgs potential in the weak field expansion of the optimal effective action~(\ref{OptimalEA}). In addition, the diffeomorphism invariance and the local SU(2) symmetry is also preserved, which is noticed at the level of the physical observables coming from the optimal scale.

After this, an essential step towards a phenomenologically viable model is taken by identifying the $\xi_i$ parameter region, for which SSB occurs and for which the weak gauge boson masses and couplings
of the SM are recovered.
This region is then compared to the explicit RG flow obtained from Asymptotic Safety coupled to matter.
It is found that only a limited composition of matter contributions allows for SSB. Further, we find it even possible to arrive at 
the phenomenologically viable IR flows that reproduce the SM masses and couplings.
However, this highly non-trivial achievement comes at the price of unnaturally large values
of either $G_0$, or $\Lambda_0$, or both.
This means that the hierarchy problem and the cosmological constant problem are not simultaneously solved if one uses this novel mechanism for SSB in a phenomenologically viable parameter range.

Concerning future extensions of this work, we plan to investigate
the higher order Higgs couplings, like
the triple Higgs coupling, which is necessarily and naturally introduced by our approach.

\subsection{Similarities and differences}

Firstly, it is important to reinforce that quantum effects can emerge when we: 
i) change the classical action to an effective action with scale-dependent couplings, 
ii) replace the classical coupling with scale-dependent couplings at the level of the equation of motion, and finally, 
iii) substitute the classical coupling for scale-dependent couplings at the level of the solution. 
For this mechanism, a key ingredient takes particular importance: the renormalization scale $k$. 
Depending on the formalism, a concrete form of $k$ should be considered (although scale-dependent gravity bypasses such a problem). 
Thus, the method used in this manuscript obtains an optimal scale $k_{\text{opt}}$ different from many other approaches when $k$ is always taken from other ad hoc physical considerations. 
Indeed, the VPS is a self-consistent form to obtain the renormalization scale, which is always the subject of doubt. In what follows, we will mention the similarities and differences between closely related approaches.

\begin{itemize}
\item Coleman Weinberg~\cite{Coleman:1973jx}: \\
This seminal paper is similar in the sense that it uses quantum corrections and scale settings to generate SSB.
It is, however, different in that the scale setting in their approach is not self-consistent in the previous sense.
Another difference is that no gravitational couplings are considered in the Coleman Weinberg approach.
\item Gravitational Higgs mechanism \cite{Bonifacio:2019mgk}: 
It is similar in the sense that it considers the gravitational sector and its potential for SSB.
However, it does not give a phenomenologically viable model for the masses of the SM
since the SSB induced by curvature itself is by far too weak.
\item Asymptotically safe Higgs \cite{Calmet:2010ze}: 
This work is similar since it considers both scale-dependence and gravitational effects.
The main difference between this and our approach is that in~\cite{Calmet:2010ze} the Higgs field is not dynamical,
an option that has been excluded since the discovery of the Higgs particle.
\end{itemize}

\subsection{Conclusion}

To conclude, let us reinforce that the main goal of the current paper 
is to implement a phenomenologically viable Higgs-like SSB, 
buy bypassing the usual quartic coupling 
and also the necessity of new particle sectors. 
This is achieved by considering: i) the gravitational sector and
ii) quantum-induced scale-dependence. 
It is shown that this possibility comes with a 
tight restriction in the parameter space of the IR-expansion
of the gravitational couplings $\xi_{\tilde{\Lambda},1},\xi_{\tilde{\Lambda},2}$, shown in figures~\ref{fig:Intersection} and~\ref{regions2}.
This finding is generic and can, in principle, be compared to 
any effective theory of quantum gravity.

Finally, as a benchmark, we study  Asymptotic Safety coupled to matter.
By contrasting the restricted $\xi_i$ parameter space to the
predictions from AS, we find
that it translates to a restriction of the particle content of the matter
sector  (see figure~\ref{ComparisonAS}) and to conditions
on the gravitational IR sector (see table~\ref{Tab.label2} and figure~\ref{RGfromAS}).

\section{Acknowledgements}

The authors are grateful to Marcelo Loewe for useful discussions. The work of C.L. is funded by Becas Chile, ANID-PCHA/2020-72210073. A. R. thanks to the Universidad de Tarapacá for support.

\newpage
\section{Appendix A: Fixed points and relevant directions}\label{AppendixA}

\begin{table}[!hb]
\begin{footnotesize}
\begin{center}
\begin{tabular}{ |c|c|c|c|c|c|  }
 \hline
 \hline
 $N_{s}$ & $N_{D}$ & $g^*$ & $\lambda^*$ & $\theta_{1}$ & $\theta_{2}$\\
 \hline
 \hline
4  & 0.0 & 0.540742 & 0.176869 & 3.23814 & 2.27473 \\ \hline 
4  & 0.5 & 0.565043 & 0.153028 & 3.2317 & 2.25416 \\ \hline 
4  & 1.0 & 0.592509 & 0.127241 & 3.28397 & 2.17999 \\ \hline 
4  & 1.5 & 0.623117 & 0.099151 & 3.35748 & 2.08812 \\ \hline 
4  & 2.0 & 0.540261 & 0.162249 & 3.14709 & 2.32315 \\ \hline 
4  & 2.5 & 0.694209 & 0.034573 & 3.50611 & 1.90999 \\ \hline 
4  & 3.0 & 0.735102 & -0.002737 & 3.57142 & 1.83258 \\ \hline 
4  & 3.5 & 0.779942 & -0.04402 & 3.62927 & 1.76401 \\ \hline
\hline
\hline 
0  & 0 & 0.540185 & 0.118598 & 3.14891 & 2.25871 \\ \hline
5  & 0 & 0.542103 & 0.191518 & 3.39657 & 2.17699 \\ \hline
10 & 0 & 0.726928 & 0.27161 & 7.51815 & 0.4841 \\ \hline
15 & 0 & 5.130506 & 0.207656 & 68.106 & 13.9233 \\ \hline
20 & 0 & 0 & 0 & 2 & -2 \\ \hline
25 & 0 & 6.214015 & 2.529234 & 3.44841 & 3.44841 \\ \hline
30 & 0 & 9.6038 & 5.23811 & 3.48826 & 3.48826 \\ \hline
35 & 0 & 39.23227 & 26.154406 & 3.81651 & 3.02378 \\ \hline
\hline
\hline
0  & 1 & 0.590545 & 0.065191 & 3.35241 & 2.0528 \\ \hline 
5  & 1 & 0.592631 & 0.142912 & 3.31242 & 2.18789 \\ \hline 
10 & 1 & 0.614441 & 0.220853 & 4.42226 & 1.65896 \\ \hline 
15 & 1 & 2.14229 & 0.241764 & 22.8232 & 4.56566 \\ \hline 
20 & 1 & 13.85667 & 0.109377 & 197.961 & 10.288 \\ \hline 
25 & 1 & 8.180358 & 2.533882 & 3.32345 & 3.32345 \\ \hline 
\hline
\hline
5  & 2 & 0.65775 & 0.085304 & 3.42891 & 2.02211 \\ \hline
10 & 2 & 0.662973 & 0.169823 & 3.81476 & 1.96256 \\ \hline
15 & 2 & 0.862502 & 0.247332 & 7.61658 & 1.01664 \\ \hline
20 & 2 & 4.654184 & 0.179771 & 52.1339 & 6.06412 \\ \hline
25 & 2 & 0 & 0 & 2 & -2 \\ \hline
\hline 
\hline 
5  & 3 & 0.738232 & 0.015283 & 3.56363 & 1.85028 \\ \hline 
10 & 3 & 0.745478 & 0.1082 & 3.68564 & 1.92326 \\ \hline 
15 & 3 & 0.786123 & 0.196828 & 5.02367 & 1.58008 \\ \hline 
20 & 3 & 2.028003 & 0.216749 & 18.8795 & 2.50838 \\ \hline 
25 & 3 & 0 & 0 & 2 & -2 \\ \hline
\hline 
\hline
10 & 4 & 0.860439 & 0.02925 & 3.72694 & 1.76028 \\ \hline
15 & 4 & 0.876891 & 0.131965 & 4.34151 & 1.71489 \\ \hline
20 & 4 & 1.165944 & 0.208745 & 8.52409 & 1.28589 \\ \hline
\hline 
\hline
15 & 5 & 1.048277 & 0.042209 & 4.13227 & 1.5924 \\ \hline 
20 & 5 & 1.149049 & 0.147856 & 6.00079 & 1.38829 \\ \hline
\hline
\hline
\end{tabular}
\end{center}
\caption{
Selected gravitational fixed points and relevant directions for different values of the set $\{N_s, N_D\}$ taking $N_V = 4$, for type II cutoff, Feynman-de Donder gauge and one loop approximation. The first and second column indicate the matter content. The third and fourth column are the fixed points for the Newtons' and cosmological constant. The fifth and sixth column represents the negative value of the critical exponents.}
\label{ta:morse}
\end{footnotesize}
\end{table}

\newpage

\section{Appendix B: Proof of U(1) gauge invariance of optimal action}\label{AppendixB}

This appendix aims to demonstrate in detail the invariance under U(1) group transformations of the optimal effective action, 
%
\begin{align}\label{OptActionU1}
    \Gamma_{\text{opt}}=&\int d^4x\sqrt{-g}\Big[2\Lambda_0+m^2_0|\phi|^2+A_\alpha A^\alpha|\phi|^2+iA^\alpha\left(\phi\partial_\alpha\phi^*-\phi^*\partial_\alpha\phi\right)+\partial_\alpha\phi^*\partial^\alpha\phi-\frac{1}{4e^2_0}F^{\mu\nu}F_{\mu\nu}\nonumber\\
    &+\frac{k^2_{\text{opt}}}{2}\Big\{4\xi_{\tilde{\Lambda},2}+2\xi_{m,2}|\phi|^2+4\xi_{\tilde{\Lambda},1}\left(8\xi_{\tilde{\Lambda},2}+4\xi_{m,2}|\phi|^2-\xi_{e,2}F^{\mu\nu}F_{\mu\nu}\right)-\frac{\xi_{e,2}}{2}F^{\mu\nu}F_{\mu\nu}\nonumber\\
    &+\xi_{e,1}\left(4\xi_{\tilde{\Lambda},2}+2\xi_{m,2}|\phi|^2-\xi_{e,2}F^{\mu\nu}F_{\mu\nu}\right)+4\xi_{m,1}|\phi|^2\left(4\xi_{\tilde{\Lambda},2}+2\xi_{m,2}|\phi|^2-\xi_{e,2}F^{\mu\nu}F_{\mu\nu}\right)\Big\}\Big].
\end{align}
An element of the \text{U(1)} group is given by
\begin{equation}
    U(x)=e^{-i\alpha(x)}
\end{equation}
the above transformation modifies the fields such that,
\begin{equation}\label{GaugeTransformationsU1}
    \phi'=U\phi, \;\;\;\;\;\;\;\;\; A_\mu'=A_\mu+\partial_\mu\alpha.
\end{equation}
These transformations imply the invariance of the following terms
\begin{align}
    |\phi|^2=\phi^*\phi \;\;\;\; \longrightarrow \;\;\;\; \phi^*U^*U\phi=\phi^*\phi=|\phi|^2,
\end{align}
\begin{align}
    F_{\mu\nu}=\partial_\mu A_\nu-\partial_\nu A_\mu \;\;\;\; \longrightarrow \;\;\;\; &\partial_\mu(A_\nu+\partial_\nu\alpha)-\partial_\nu(A_\mu+\partial_\mu\alpha)\nonumber\\
    =&\partial_\mu A_\nu+\partial_\mu\partial_\nu\alpha-\partial_\nu A_\mu-\partial_\nu\partial_\mu\alpha\nonumber\\
    =&\partial_\mu A_\nu-\partial_\nu A_\mu\nonumber\\
    =&F_{\mu\nu}.
\end{align}
Therefore, the optimal scale
\begin{equation}
    k_{\text{opt}}=-\frac{8\xi_{\tilde{\Lambda},1}+4\xi_{m,1}|\phi|^2 -\xi_{e,1}F^{\mu\nu}F_{\mu\nu}}{2\left(8\xi_{\tilde{\Lambda},2}+4\xi_{m,2}|\phi|^2-\xi_{e,2}F^{\mu\nu}F_{\mu\nu}\right)},
\end{equation}
is invariant under the U(1) gauge transformations~(\ref{GaugeTransformationsU1}). Once having proved the invariance of the quadratic self-interaction $|\phi|^2$ and the kinetic term for the gauge field $F^{\mu\nu} \, F_{\mu\nu}$, the only remaining terms where the invariance has to be demonstrated are those involving non-minimal interactions between the gauge and scalar field, gauge field self-interactions and the kinetic term for the scalar $\phi$ present in the first line of~(\ref{OptActionU1}). By applying~(\ref{GaugeTransformationsU1}), one obtains,

\begin{align}
    &A_\mu A^\mu|\phi|^2+iA^\mu\phi\partial_\mu\phi^*-iA^\mu\phi^*\partial_\mu\phi+\partial_\mu\phi^*\partial^\mu\phi\nonumber\\
    \longrightarrow \;\;\;\; &(A_\mu+\partial_\mu\alpha)(A^\mu+\partial^\mu\alpha)|\phi|^2+i(A^\mu+\partial^\mu\alpha)U\phi\partial_\mu(\phi^*U^*)-i(A^\mu+\partial^\mu\alpha)\phi^*U^*\partial_\mu(U\phi)+\partial_\mu(\phi^*U^*)\partial^\mu(U\phi)\nonumber\\
    =&A_\mu A^\mu |\phi|^2+A_\mu\partial^\mu\alpha |\phi|^2+\partial_\mu\alpha A^\mu |\phi|^2+\partial_\mu\alpha\partial^\mu\alpha |\phi|^2+iA^\mu U\phi\partial_\mu(\phi^*U^*)+i\partial^\mu\alpha U\phi\partial_\mu(\phi^*U^*)\nonumber\\
    &-iA^\mu\phi^*U^*\partial_\mu(U\phi)-i\partial^\mu\alpha\phi^*U^*\partial_\mu(U\phi)+(\partial_\mu\phi^*U^*+i\phi^*\partial_\mu\alpha U^*)(U\partial^\mu\phi-iU\phi\partial_\mu\alpha)\nonumber\\
    =&A_\mu A^\mu |\phi|^2+A_\mu\partial^\mu\alpha |\phi|^2+\partial_\mu\alpha A^\mu |\phi|^2+\partial_\mu\alpha\partial^\mu\alpha |\phi|^2+iA^\mu\phi\partial_\mu\phi^*-A^\mu|\phi|^2\partial_\mu\alpha+i\partial^\mu\alpha\phi\partial_\mu\phi^*\nonumber\\
    &-\partial^\mu\alpha\partial_\mu\alpha|\phi|^2-A^\mu|\phi|^2\partial_\mu\alpha-iA^\mu\phi^*\partial_\mu\phi-\partial^\mu\alpha\partial_\mu\alpha|\phi|^2-i\partial^\mu\alpha\phi^*\partial_\mu\phi+\partial_\mu\phi^*\partial^\mu\phi-i\partial_\mu\phi^*\phi\partial^\mu\alpha\nonumber\\
    &+i\phi^*\partial_\mu\alpha\partial^\mu\phi+\partial_\mu\alpha\partial^\mu\alpha|\phi|^2\nonumber\\
    =&A_\mu A^\mu|\phi|^2+iA^\mu\phi\partial_\mu\phi^*-iA^\mu\phi^*\partial_\mu\phi+\partial_\mu\phi^*\partial^\mu\phi
\end{align}

Therefore, the optimal action is completely invariant under \text{U(1)} transformation.
The same is true for the \text{SU(2)} case. The optimal action is invariant under transformations of the group \text{SU(2)}.



\end{document}